\documentclass[12pt]{article}

\usepackage[a4paper,
            twoside = true,
			top = 30mm,
		    bottom = 30mm,
            inner = 25mm,
            outer = 25mm
		    ]{geometry}
            
\usepackage{amsmath,amsfonts,mathtools,braket,amssymb,slashed,euscript,bm,bbm,graphicx}
\usepackage[table,dvipsnames]{xcolor}
\usepackage[labelfont=bf,width=0.95\textwidth]{caption}
\usepackage{cite}
\usepackage{cancel}
\usepackage[
    bookmarksnumbered=true,
    urlcolor=blue,
    linkbordercolor=red,
    citebordercolor=green,
    bookmarksopen=true
    ]{hyperref}

\numberwithin{equation}{section}   
\allowdisplaybreaks                 

\makeatletter
\DeclareRobustCommand{\cev}[1]{%
  {\mathpalette\do@cev{#1}}%
}
\newcommand{\do@cev}[2]{%
  \vbox{\offinterlineskip
    \sbox\z@{$\m@th#1 x$}%
    \ialign{##\cr
      \hidewidth\reflectbox{$\m@th#1\vec{}\mkern4mu$}\hidewidth\cr
      \noalign{\kern-\ht\z@}
      $\m@th#1#2$\cr
    }%
  }%
}
\makeatother

\newcommand{\braces}[1]{[\hspace{-0.5mm}[ #1 ]\hspace{-0.5mm}]}
\newcommand{\Braces}[1]{\big[\hspace{-1mm}\big[ #1 \big]\hspace{-1mm}\big]}

\newcommand{\spac}{{\hspace{0.3mm}}}
\newcommand{\np}{n_+}
\newcommand{\nm}{n_-}
\newcommand{\nps}{\spac/\hspace{-2.3mm}n_+}
\newcommand{\nms}{\spac/\hspace{-2.3mm}n_-}
\newcommand{\Ch}{\chi}
\newcommand{\A}{\mathcal{A}}
\newcommand{\As}{\hspace{1mm}/\hspace{-3mm}\A}
\newcommand{\C}{{c}}
\newcommand{\Cb}{{\bar{c}}}

\newcommand{\lnQ}{L}
\newcommand{\lnsqQ}{L^2}
\newcommand{\Lx}{L_x}
\newcommand{\Lxb}{L_{\bar{x}}}
\newcommand{\Lsqx}{L_x^2}
\newcommand{\Lsqxb}{L_{\bar{x}}^2}

\allowdisplaybreaks

\title{One-loop matching of QCD currents to \texorpdfstring{\\[2mm]}{ } power-suppressed two-jet operators}

\begin{document}

\begin{titlepage}

\begin{flushright}
{\small
TUM-HEP-1606/26\\
July~1, 2026
}
\end{flushright}

\makeatletter
\vskip0.8cm
\pdfbookmark[0]{\@title}{title}
\begin{center}
{\Large\bf\boldmath \@title}
\end{center}
\makeatother

\vspace{0.5cm}
\begin{center}
\textsc{Martin~Beneke}, 
\textsc{Aleksey~V.~Rusov},
and \textsc{Michel~Stillger}\\[6mm]
\textit{Physik Department T31, Technische Universit\"at M\"unchen\\
James-Franck-Straße 1, D-85748 Garching, Germany}
\end{center}

\vspace{0.6cm}
\pdfbookmark[1]{Abstract}{abstract}
\begin{abstract}
\noindent 
We compute the matching of QCD quark-antiquark currents onto the set of the two-particle and three-particle two-jet operators in soft-collinear effective theory (SCET) at next-to-leading order (NLO) in the perturbative QCD series, 
including for the first time operators up to second order in the power expansion 
in the transverse momentum over energy. These results contribute to the ongoing 
programme of computing power corrections and summing power-suppressed logarithmically enhanced terms for event shapes in the two-jet region and deep-inelastic scattering 
in the Bjorken-$x\to 1$ limit. The three-particle operators depend on the partonic momentum fractions of two partons moving into the same direction. When one of the momentum fractions approaches zero, the coefficient functions are shown to satisfy endpoint factorization relations, which allow for a consistent cancellation of endpoint singularities among various terms in the complete factorization formula for power corrections.   
\end{abstract}

\end{titlepage}

\section{Introduction}
\label{sec:intro}

In many hard processes, extending theoretical predictions beyond the leading-power approximation in the high-energy expansion and systematically incorporating subleading contributions is already relevant to match the accuracy of experimental data (see, for example, refs.~\cite{MovillaFernandez:2001ed,Nason:2023asn,Benitez:2024nav} for event shapes, refs.~\cite{Beneke:2019mua,vanBeekveld:2021hhv,AH:2021kvg} for Drell-Yan production, chapter~10 of ref.~\cite{Boussarie:2023izj} for transverse-momentum dependent processes). 
Within this context, factorization provides a powerful framework for separating short- and long-distance dynamics.
In many high-energy reactions, particles and jets carry large energies while their invariant masses remain parametrically small.
This scale hierarchy naturally suggests a description in terms of soft-collinear effective theory (SCET)~\cite{Bauer:2000yr,Bauer:2001yt,Beneke:2002ph,Beneke:2002ni}, which systematically exploits the separation between hard, collinear and soft degrees of freedom.

In SCET, one matches full theory operators (in many applications currents in QCD) onto a basis of leading- and subleading-power ``$N$-jet operators'' consisting of products of collinear fields describing collections of partons emerging from the hard process, moving into $N$ widely separated directions. The short-distance coefficients of these  operators encode the hard physics. 
The most prominent example is the electromagnetic current $\bar\psi \gamma^\mu \psi$, which is sourced by a virtual photon.
Similar considerations apply to axial-vector, scalar and pseudoscalar currents, which appear in a variety of phenomenological settings and exhibit an analogous matching structure.

\begin{figure}[t]
    \centering
    \includegraphics[scale=0.5]{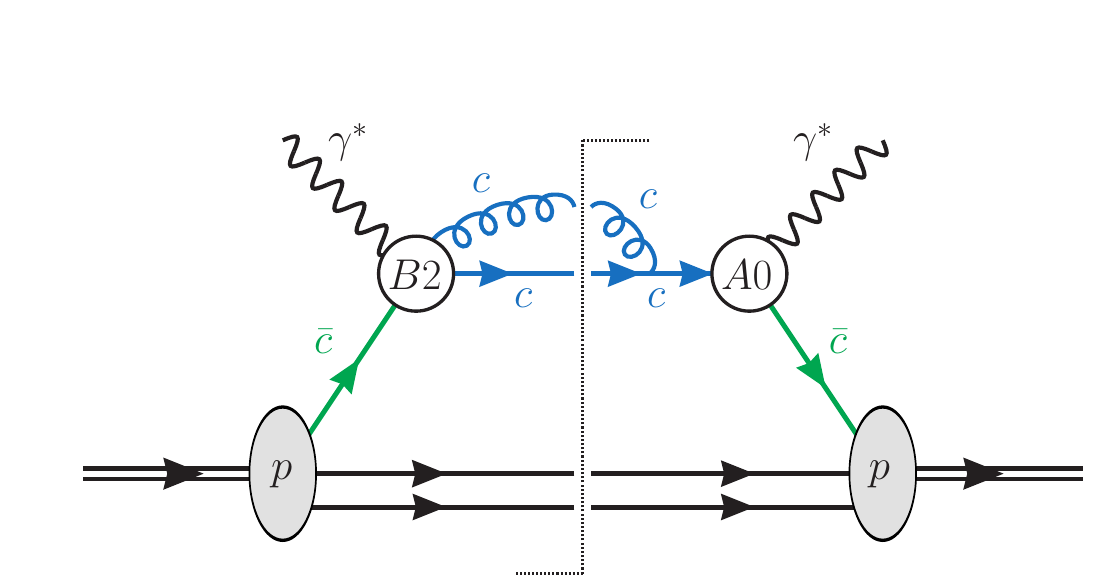}
    \caption{Contribution to the next-to-leading factorization theorem for DIS involving hard operators with three fields.}
    \label{fig:DIS}
\end{figure}

We consider the most important case of $N=2$, in which it is convenient to adopt the reference frame, in which the two directions, referred to as "collinear"~($\C$) and "anti-collinear"~($\Cb$) are nearly back-to-back.
Examples are presented by the production of two-jets from a non-hadronic initial state, or deep-inelastic scattering (DIS) in the Bjorken-$x\to 1$ region when the final state is kinematically squeezed into a single jet.
The characteristic feature of such processes is that the transverse momentum $p_\perp$ of particles relative to the back-to-back axis is much smaller than the hard scale $Q$, set by the energy of the jets.
The expansion of the hard process in $1/Q$ is therefore described by operators with an increasing number of transverse derivatives or an increasing number of fields that emanate from the hard process in a given direction. Schematically
\begin{align}
    \lambda^0: &\qquad \Ch_\C
    \nonumber\\*
    \lambda^1: &\qquad \partial_\perp\Ch_\C\,,\;\A_{\C\perp}\Ch_\C
    \nonumber\\*
    \lambda^2: &\qquad \partial_\perp\partial_\perp\Ch_\C\,,\;\big[\partial_\perp\A_{\C\perp}\big]\Ch_\C\,,\; \A_{\C\perp} \big[\partial_\perp\Ch_\C\big]\,,\; \dots  
\end{align}
where $\lambda=p_\perp/Q$ is the dimensionless power-counting parameter.
The above list refers to jets, which are sourced to leading power by quarks.
We have given power-suppressed operators up to order $\lambda^2$, which describe single jets formed from up to two partons at the hard vertex.
The ellipsis denote operators with three fields and no transverse derivative, which also contribute at $\mathcal{O}(\lambda^2)$.
The renormalization of these operators has been computed at the one-loop order \cite{Beneke:2017ztn, Beneke:2018rbh}, but the matching coefficients of two-jet operators for QCD currents are known only at leading power~\cite{Bauer:2003di} 
(to four loops from ref.~\cite{Lee:2022nhh}), and at subleading power 
$\mathcal{O}(\lambda)$ for heavy-light currents \cite{Beneke:2004rc,Hill:2004if} and back-to-back jets \cite{Strohm:2020,Vladimirov:2021hdn}. The present work reports the missing one-loop matching coefficients for the  $\mathcal{O}(\lambda^2)$ suppressed operators in the third line above. In processes that do not resolve azimuthal dependence, the $\mathcal{O}(\lambda)$ correction is absent and these $\mathcal{O}(\lambda^2)$ coefficients contribute to the same next-to-leading power (NLP) term as double insertions of the  $\mathcal{O}(\lambda)$ operators in the second line. 
As an example, in fig.~\ref{fig:DIS} a contribution to the NLP factorization theorem for large-$x$ DIS involving subleading-power SCET operators is shown. The nomenclature ``B2" refers to 
$\left[\partial_\perp\A_{\C\perp}\right]\Ch_\C\,,
   \A_{\C\perp} \left[\partial_\perp\Ch_\C\right]$ and will be explained below.

This paper is structured at follows:
First, we give the relevant SCET definitions and operators in sec.~\ref{sec:basis} before describing the matching procedure and presenting our main results in sec.~\ref{sec:matching_coefficients}.
We then provide two consistency checks of our results by  comparing to known results for the anomalous dimensions of subleading-power operators in~sec.~\ref{sec:renormalization}, and by deriving the factorization of the matching coefficients in the soft-gluon limit in~sec.~\ref{sec:refactorization}.
We conclude in sec.~\ref{sec:conclusion}.
While the main text is focused on the (axial-)vector current, our results are extended in appendix~\ref{app:scalar-current} to (pseudo)scalar currents. Finally, in appendices~\ref{app:MoR-trick}~to~\ref{app:bases-conv},  we present some supplementary material on the analysis performed.


\section{Operator basis}
\label{sec:basis}

As the particles carry high energies $Q$ but have small invariant mass, their four-momenta are light-like, and it is convenient to decompose momenta in a light-cone basis
\begin{align}
    p^\mu &= \np p \, \frac{\nm^\mu}{2} + \nm p \, \frac{\np^\mu}{2} + p_\perp^\mu \equiv (\np p, \nm p, p_\perp^\mu) \,,  
\end{align}
where $n_\pm^2=0$ and $\np\nm=2$.
In SCET, one then introduces separate fields to describe the relevant momentum modes.
In many applications, one considers back-to-back light-like directions.
The relevant modes are then collinear $p_\C\sim(1,\lambda^2,\lambda)Q$, anti-collinear $p_\Cb\sim(\lambda^2,1,\lambda)Q$, and soft $p_s\sim(\lambda^2,\lambda^2,\lambda^2)Q$. The effective theory is called SCET${}_\mathrm{I}$ in this case.
The power-counting parameter $\lambda\ll1$ is set by a small momentum scale in the problem under consideration.
In the SCET construction, hard modes $p_h\sim(1,1,1)Q$ are integrated out, and their effect is captured in short-distance coefficients.

The electro-magnetic current operator is matched on the position-space formulation~\cite{Beneke:2002ph,Beneke:2002ni} of SCET${}_\mathrm{I}$ according to 
\begin{align} \label{eq:match-rel}
    \bar\psi \spac \gamma^\mu \spac \psi(0) &= \int\!dt \, ds \, \sum_i C_i(t,s) \, J_i(t,s)
    \\* \nonumber
    &+ \int\!dt \, d^2s \, \sum_i C_i(t,\{\underline{s}\}) \, J_i(t,\{\underline{s}\}) + \int\!d^2t \, ds \, \sum_i C_i(\{\underline{t}\},s) \, J_i(\{\underline{t}\},s) + \dots \,, 
\end{align}
where ellipsis denote terms with more than three fields, $d^2s\equiv ds_1ds_2$ and $\{\underline{s}\}\equiv\{s_1,s_2\}$.
Here and in the following, we suppress the Lorentz index $\mu$ on the SCET operators $J_i$ to increase readability.
Both the operators and the short-distance coefficients $C_i$ are functions of the variables $t,s$, which describe the displacement of the SCET fields along the $\nm$ and $\np$ light cones, respectively.
This non-locality arises in SCET because derivatives $\np\partial$ ($\nm\partial$) acting on (anti-)collinear fields are not power-suppressed.
In momentum space this corresponds to the fact that hard subgraphs cannot be expanded in the large component $\np p$ ($\nm p$) of an external (anti-)collinear momentum relative to the hard scale $Q$, so that the momentum-space short-distance coefficients $C_i$ are functions of these $\mathcal{O}(1)$ quantities. 

The SCET${}_\mathrm{I}$ operators $J_i$ on the right-hand side of eq.~\eqref{eq:match-rel} are constructed in terms of the gauge-invariant building blocks $\Ch_\C$, $\A_{\C\perp}$ and $\Ch_\Cb$, $\A_{\Cb\perp}$ as well as transverse derivatives acting on those~\cite{Beneke:2017ztn}.
These objects are all of $\mathcal{O}(\lambda)$ in power counting.
The relevant definitions are
\begin{align} \label{eq:BBs}
    \Ch_\C &\equiv W_\C^\dagger \, \xi_\C \,,
    &
    \A_\C^\mu &\equiv W_\C^\dagger \, g_s \spac A_\C^\mu \spac W_\C + W_\C^\dagger \, [iD_s^\mu, W_\C] \,,  
\end{align}
where $\xi_\C$ and $A_\C$ are fields describing collinear modes only, which satisfy $\nms\xi_\C=0$ and $\np\A_\C=0$.
The bare strong coupling is denoted by $g_s$ and $iD_s^\mu=i\partial^\mu+g_s \spac \nm A_s \, \np^\mu/2$ is the covariant derivative containing the soft background field.
The Wilson lines
\begin{align}
    W_\C (x) = \text{P} \exp\!\bigg[i g_s \int_{-\infty}^{0}\!ds \, \np A_\C(x + s \np) \bigg]  
\end{align}
ensure gauge invariance and contain the unsuppressed component of the collinear gluon field.
The definitions in the anti-collinear sector can be obtained by $\C\leftrightarrow\Cb$ and $\np\leftrightarrow\nm$.
The EFT operators are constructed such that they have mass dimension 3 and are boost invariant, i.e.\ invariant under $\np\to\zeta\np$ and $\nm\to\zeta^{-1}\nm$. Besides this, one uses the projection properties $\nms\Ch_\C=\nps\Ch_\Cb=0$ to reduce the number of Dirac structures.

The labeling of the basis operators $J^{(Xn,Ym)}$ with $X,Y\in\{A,B,C,\dots\}$ and $n,m\in\{0,1,2,\dots\}$ below follows the notation in~refs.~\cite{Beneke:2004rc,Beneke:2017ztn,Beneke:2018rbh}.
The first entry in parenthesis refers to the anti-collinear sector, whereas the second refers to the collinear one.
The number of (anti-)collinear fields is denoted by the capital letters, with $A$ referring to one, $B$ to two building blocks, and so on.
The number indicates the power suppression in this sector relative to the leading-power operator~$(A0)$ which contains only a single (anti-)collinear field.

At leading power in the $\lambda$ expansion, there exists only the single operator
\begin{align} \label{eq:J-A0}
    J^{(A0,A0)}(t,s) = \bar\Ch_\Cb (t\nm) \, \gamma_\perp^\mu \, \Ch_\C (s\np) \,.  
\end{align}
The $\mathcal{O}(\lambda)$ power suppression can arise in two different ways.
Either one includes transverse derivatives acting on the gauge-invariant building blocks
\begin{align} \label{eq:J-A1}
    J_1^{(A0,A1)}(t,s) &= \bar\Ch_\Cb(t\nm) \, \frac{\np^\mu}{i \np \partial} \,  i \slashed{\partial}_\perp \spac \Ch_\C (s\np) \,,  
    \nonumber\\
    J_2^{(A0,A1)}(t,s) &= \bar\Ch_\Cb(t\nm) \, \frac{\nm^\mu}{i \nm \cev{\partial}} \, i \slashed{\partial}_\perp \spac \Ch_\C (s\np) \,,  
\end{align}
or an additional gluon building block 
\begin{align} \label{eq:J-B1}
    J_1^{(A0,B1)}(t,s_1, s_2) &= \bar\Ch_\Cb(t\nm) \, \frac{\np^\mu}{i \np \partial} \, \As_{\C\perp}(s_1\np) \, \Ch_\C (s_2\np) \,,  
    \nonumber\\
    J_2^{(A0,B1)}(t,s_1, s_2) &= \bar\Ch_\Cb(t\nm) \, \frac{\nm^\mu}{i \nm  \cev{\partial}} \, \As_{\C\perp}(s_1\np) \, \Ch_\C (s_2\np) \,,  
\end{align}
where the inverse derivative $(i\np\partial)^{-1}$ acts on \emph{all} fields to its right.
The operators $J_i^{(A1,A0)}$ and $J_i^{(B1,A0)}$ can be obtained by the replacements $\C\leftrightarrow\Cb$ and $\np\leftrightarrow\nm$.

For the $\mathcal{O}(\lambda^2)$ operators, it is possible to combine power suppression from transverse derivatives and additional gluon building blocks.
One finds six operators without additional fields. We choose them to be
\begin{align} \label{eq:J-A1A1}
    J_1^{(A1,A1)}(t,s) &= \bar\Ch_\Cb(t\nm) \, \frac{i \cev{\slashed{\partial}}_\perp \spac \gamma_\perp^\mu \spac i \slashed{\partial}_\perp}{i \nm \cev{\partial} \, i \np \partial} \, \Ch_\C (s\np) \,,  
    \nonumber\\
    J_2^{(A1,A1)}(t,s) &= \bar\Ch_\Cb (t\nm) \, \frac{i \cev{\partial}_\perp^\nu \spac \gamma_\perp^\mu \spac i\partial_{\perp\nu}}{i \nm \cev{\partial} \, i \np \partial} \, \Ch_\C (s\np) \,,  
    \nonumber\\
    J_3^{(A1,A1)}(t,s) &= \bar\Ch_\Cb (t\nm) \, \frac{i \cev{\slashed{\partial}}_\perp \spac i\partial_\perp^\mu}{i \nm \cev{\partial} \, i \np \partial} \, \Ch_\C (s\np) \,,  
    \nonumber\\
    J_4^{(A1,A1)}(t,s) &= \bar\Ch_\Cb (t\nm) \, \frac{i \cev{\partial}_\perp^\mu \spac i \slashed{\partial}_\perp}{i \nm \cev{\partial} \, i \np \partial} \, \Ch_\C (s\np) \,,  
\end{align}
and
\begin{align} \label{eq:J-A2}
    J_1^{(A0,A2)}(t,s) &= \bar\Ch_\Cb (t\nm) \, \frac{\gamma_\perp^\mu \, (i\partial_\perp)^2}{i \nm \cev{\partial} \, i \np \partial} \, \Ch_\C (s\np) \,,  
    \nonumber\\
    J_2^{(A0,A2)}(t,s) &= \bar\Ch_\Cb (t\nm) \, \frac{i\partial_\perp^\mu \, i\slashed{\partial}_\perp}{i \nm \cev{\partial} \, i \np \partial} \, \Ch_\C (s\np) \,.  
\end{align}
In total there are twelve operators with an additional gluon building block at $\mathcal{O}(\lambda^2)$.
Four of those contain a transverse derivative acting on the anti-collinear quark field
\begin{align} \label{eq:J-A1B1}
    J_1^{(A1,B1)}(t,s_1,s_2) &= \bar\Ch_\Cb (t\nm) \, \frac{i \cev{\slashed{\partial}}_\perp \spac \gamma_\perp^\mu}{i \nm\cev{\partial} \, i\np\partial} \, \As_{\C\perp}(s_1\np) \, \Ch_\C (s_2\np) \,,  
    \nonumber\\
    J_2^{(A1,B1)}(t,s_1,s_2) &= \bar\Ch_\Cb (t\nm) \, \frac{i \cev{\slashed{\partial}}_\perp}{i \nm\cev{\partial} \, i \np\partial} \, \A_{\C\perp}^\mu(s_1\np) \, \Ch_\C (s_2\np) \,,  
    \nonumber\\
    J_3^{(A1,B1)}(t,s_1,s_2) &= \bar\Ch_\Cb (t\nm) \, \frac{i \cev{\partial}_\perp^\mu}{i \nm\cev{\partial} \, i \np\partial} \, \As_{\C\perp}(s_1\np) \, \Ch_\C (s_2\np) \,,  
    \nonumber\\
    J_4^{(A1,B1)}(t,s_1,s_2) &= \bar\Ch_\Cb (t\nm) \, \frac{i\cev{\partial}_{\perp\rho}\spac \gamma_\perp^\mu}{i \nm\cev{\partial} \, i \np\partial} \, \A_{\C\perp}^\rho (s_1 n_+) \, \Ch_\C (s_2\np) \,.  
\end{align}
The remaining eight operators contain a transverse derivative in the collinear sector.
We choose them to be
\begin{align} \label{eq:J-B2}
    J_1^{(A0,B2)}(t,s_1,s_2) &= \bar\Ch_\Cb(t\nm) \, \frac{\gamma_\perp^\mu}{i \nm\cev{\partial} \, i \np\partial} \, i\slashed{\partial}_\perp \big[ \As_{\C\perp}(s_1\np) \, \Ch_\C (s_2\np) \big] \,,  
    \nonumber\\
    J_2^{(A0,B2)}(t,s_1,s_2) &= \bar\Ch_\Cb(t\nm) \, \frac{1}{i \nm\cev{\partial} \, i \np\partial} 
    \, i\slashed{\partial}_\perp \big[\A_{\C\perp}^\mu(s_1\np) \, \Ch_\C (s_2\np) \big] \,,  
    \nonumber\\
    J_3^{(A0,B2)}(t,s_1,s_2) &= \bar\Ch_\Cb(t\nm) \, \frac{1}{i \nm\cev{\partial} \, i \np\partial} \, i\partial_\perp^\mu \big[ \As_{\C\perp}(s_1\np) \, \Ch_\C(s_2\np) \big] \,,  
    \nonumber\\
    J_4^{(A0,B2)}(t,s_1,s_2) &= 
    \bar\Ch_\Cb(t\nm) \, \frac{\gamma_\perp^\mu}{i \nm\cev{\partial} \, i \np\partial} \, i\partial_{\perp\rho} \big[ \A_{\C\perp}^\rho (s_1\np) \, \Ch_\C (s_2\np) \big] \,,  
    \nonumber\\
    J_5^{(A0,B2)}(t,s_1,s_2) &= \bar\Ch_\Cb(t\nm) \, \frac{\gamma_\perp^\mu}{i \nm\cev{\partial} \, i \np\partial} \, \As_{\C\perp} (s_1\np) \, i\slashed{\partial}_\perp \Ch_\C (s_2\np) \,,  
    \nonumber\\
    J_6^{(A0,B2)}(t,s_1,s_2) &= \bar\Ch_\Cb(t\nm) \, \frac{1}{i \nm\cev{\partial} \, i \np\partial} \, \A_{\C\perp}^\mu (s_1\np) \, i \slashed{\partial}_\perp \Ch_\C (s_2\np) \,,  
    \nonumber\\
    J_7^{(A0,B2)}(t,s_1,s_2) &= \bar\Ch_\Cb(t\nm) \, \frac{1}{i \nm\cev{\partial} \, i \np\partial} \, \As_{\C\perp} (s_1\np) \, i\partial_\perp^\mu \spac \Ch_\C (s_2\np) \,,  
    \nonumber\\
    J_8^{(A0,B2)}(t,s_1,s_2) &= \bar\Ch_\Cb(t\nm) \, \frac{\gamma_\perp^\mu}{i \nm\cev{\partial} \, i \np\partial} \, \A_{\C\perp}^\rho (s_1\np) \, i\partial_{\perp\rho} \spac \Ch_\C (s_2\np) \,,  
\end{align}
The operators $J_i^{(A2,A0)}$, $J_i^{(B1,A1)}$ and $J_i^{(B2,A0)}$ can again be obtained by replacing $\C\leftrightarrow\Cb$ and $\np\leftrightarrow\nm$.
We restrict our analysis to operators containing at most one additional gluon building block, i.e.\ we do not consider $(B1,B1)$- or $(A0,C2)$-type operators, 
which have four fields in total.

The choice of operators shown above is not unique.
On the one hand, the transverse derivative in the $J_i^{(A0,B2)}$ operators can be chosen to act on each collinear field separately and not their product.\footnote{In principle, there is a third option, the transverse derivative acts on the product and only the gluon field. However, due the product rule, these operators are trivially related to the ones mentioned in the main text. }
This basis choice is relevant for sec.~\ref{sec:refactorization} and is discussed in appendix~\ref{app:bases-conv}. On the other hand,  one can choose a different order of the three $\gamma$ matrices for the operators $J_1^{(A1,A1)}$, $J_1^{(A1,B1)}$, $J_1^{(A0,B2)}$ and $J_5^{(A0,B2)}$.

In this work, we use dimensional regularization with $d=4-2\epsilon$.
In $d=4$ dimensions, not all of the above operators are independent. One finds
\begin{align} \label{eq:evanescent_operators}
    0 &= J_1^{(A1,A1)} + J_2^{(A1,A1)} - J_3^{(A1,A1)} - J_4^{(A1,A1)} \equiv \hat{J}_1^{(A1,A1)} \,,  
    \nonumber\\
    0 &= J_1^{(A1,B1)} - J_2^{(A1,B1)} - J_3^{(A1,B1)} + J_4^{(A1,B1)} \equiv \hat{J}_1^{(A1,B1)} \,,  
    \nonumber\\
    0 &= J_1^{(A0,B2)} + J_2^{(A0,B2)} - J_3^{(A0,B2)} - J_4^{(A0,B2)} \equiv \hat{J}_1^{(A0,B2)} \,,  
    \nonumber\\
    0 &= J_5^{(A0,B2)} - J_6^{(A0,B2)} + J_7^{(A0,B2)} - J_8^{(A0,B2)} \equiv \hat{J}_5^{(A0,B2)} \,,  
\end{align}
as there exist only two transverse directions (i.e.~$\varepsilon^{\mu_\perp\nu_\perp\rho_\perp\sigma}=0$). The right-hand side of~eq.~\eqref{eq:evanescent_operators} therefore defines four evanescent operators $\hat{J}_i$.
In some applications, it may be convenient to eliminate the four operators containing three $\gamma$ matrices in favour of these evanescent operators.  
By construction, the short-distance coefficients of the evanescent operators are the same as the ones with three $\gamma$ matrices. However, the Wilson coefficients of the other operators change, as will be discussed below.
In the following, we sometimes use the shorthand notation $(X)\equiv(A0,X)$.

In massless QCD, the short-distance matching coefficients  for the QCD axial-vector current 
$\bar\psi\gamma^\mu\gamma_5\psi$ are trivially related to the ones for the vector case.
When working in naive dimensional regularization, i.e.\ with anticommuting $\gamma_5$, one can always anticommute the $\gamma_5$ to the position next to the collinear quark field $\Ch_\C$ and then finds $C_i^\text{axial-vector} = C_i^\text{vector}$.


\section{Matching coefficients}
\label{sec:matching_coefficients}

The matching coefficients $C_i$ are determined by evaluating the defining equation~\eqref{eq:match-rel} in appropriate \emph{on-shell} matrix elements.
For the $A$-type operators, which do not contain an additional gluon building block, we use $\langle \bar{q}(p) q(k)| \dots |0 \rangle$ and choose the momenta $p$ ($k$) to have (anti-)collinear scaling.
For the $B$-type operators, containing one additional gluon building block, we project with $\langle g(p_1) \bar{q}(p_2) q(k)| \dots |0 \rangle$ where both $p_1$ and $p_2$ have collinear scaling.

To determine the matching coefficients, it is sufficient to consider diagrams contributing to the matrix element of the left-hand side of~eq.~\eqref{eq:match-rel} with hard propagators only. 
There is one exception to this, which will be discussed below in detail.

\subsubsection*{\boldmath Coefficients of $A$-type operators}

Projecting the right-hand side of~eq.~\eqref{eq:match-rel} with an external quark-antiquark state, one finds
\begin{align} \label{eq:A-RHS}
    &\int\!dt \, ds \, \sum_i C_i(t,s) \, \langle \bar{q}(p) q(k)| \spac J_i(t,s) \spac |0 \rangle
    \nonumber\\*[-2mm]
    &={} C^{(A0)}(\nm k,\np p) \, \bar{u}_\Cb(k) \, \gamma_\perp^\mu \, v_\C(p) + C^{(A1)}_1(\nm k,\np p) \, \bar{u}_\Cb(k) \, \frac{\np^\mu \, \slashed{p}\vphantom{p}_\perp}{\np p} \, v_\C(p) + \dots \,,  
\end{align}
where the colour of the two quarks is kept implicit and we only show the first two terms.
The remaining terms, indicated by the ellipsis, can simply be obtained by replacing derivatives $i\partial \to-p$ ($-k$) acting on (anti-)collinear fields in the matrix elements of the operators defined in~eqs.~\eqref{eq:J-A1}, \eqref{eq:J-A1A1} and~\eqref{eq:J-A2}. 
The right-hand side does not receive loop corrections, as all (anti-)collinear integrals are scaleless when working on-shell.
The Fourier transformations of the $A$-type short-distance coefficients are defined by
\begin{align} \label{eq:CA-mom}
    C_i(\nm k,\np p) = \int\!dt \, ds \, C_i(t,s) \, e^{i\spac t \spac \nm k+i \spac s \spac \np p} \,.  
\end{align}
Due to boost invariance, the momentum-space coefficients are functions of $\nm k\,\np p\equiv Q^2$ only.
The SCET spinors satisfy the projection properties $\nms v_\C = \nps u_\Cb = 0$ similar to the fields.

\begin{figure}[t]
    \centering
    \begin{tabular}{ccc}
        \includegraphics[scale=0.48]{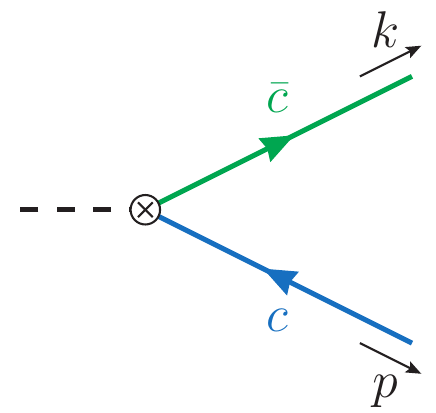} &\hspace*{12mm}& \\[-3.22cm]
        &&\includegraphics[scale=0.48]{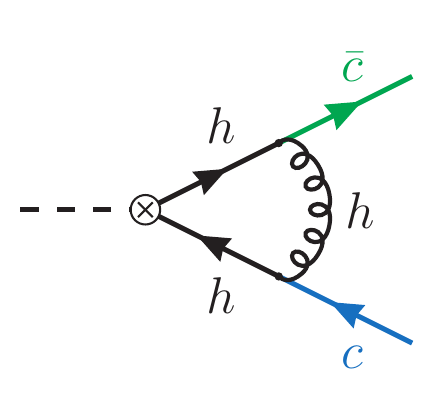} 
        \\[-0.2cm]
        (a) & & (b) 
    \end{tabular}
    \caption{Tree and one-loop diagrams contributing to the left-hand side of~eq.~\eqref{eq:match-rel} when projected with the matrix element $\langle \bar{q}(p) q(k)| \dots |0 \rangle$. Blue (green) lines represent (anti-)-collinear ($c$, resp. $\bar{c}$) fields and black ones hard ($h$) propagators. }
    \label{fig:A-diagrams}
\end{figure}

The matrix element on the left-hand side of eq.~\eqref{eq:A-RHS} is calculated in full QCD and then expanded in $\lambda$.
The QCD spinors are related to the SCET ones by~\cite{Beneke:2002ph}
\begin{align} \label{eq:QCD-SCET-spinor}
    v(p) &= \bigg( 1 - \frac{\nps}{2} \, \frac{\slashed{p}\vphantom{p}_\perp}{\np p} \bigg) \, v_\C(p) \,,
    &
    \bar{u}(k) &= \bar{u}_\Cb(k) \, \bigg( 1 - \frac{\slashed{k}_\perp}{\nm k} \, \frac{\nms}{2} \bigg) \,,  
\end{align}
where the second term in the parenthesis are of $\mathcal{O}(\lambda)$ and arise from integrating out the small components of the quark field in the SCET construction.
The on-shell conditions $p^2=k^2=0$ are used to eliminate the small momentum components $\np k,\nm p\sim\lambda^2Q$ of the (anti-)collinear momenta.
For loop diagrams, it is important to expand the integrands in $\lambda$ prior to performing the integral~\cite{Beneke:1997zp} to pick up the hard region only.
In appendix~\ref{app:MoR-trick} a simplification to this method is discussed.
Calculating the two diagrams in fig.~\ref{fig:A-diagrams} 
in Feynman gauge using the \texttt{Mathematica} packages \textsc{FeynCalc}~\cite{Shtabovenko:2023idz} and \textit{package}-\textbf{X}~\cite{Patel:2015tea}, one finds by comparing to~eq.~\eqref{eq:A-RHS} the short-distance coefficients of the $A$-type operators.
It turns out that all non-zero subleading coefficients are related to~\cite{Bauer:2003di}
\begin{align} \label{eq:CA0}
    C^{(A0)} &= 1 + \frac{\alpha_s}{4\pi} \, C_F \bigg[-\frac{2}{\epsilon^2} + \frac{2\spac \lnQ -3}{\epsilon} - \lnsqQ + 3 \spac \lnQ - 8 + \frac{\pi^2}{6} \bigg] + \mathcal{O}(\alpha_s^2) \,,  
\end{align}
where $\alpha_s\equiv\alpha_s(\mu)$ is the in the $\overline{\text{MS}}$ scheme renormalized strong coupling constant, and throughout this work we adopt the following shorthand notations for the logarithms
\begin{align} \label{eq:logs}
    \lnQ &\equiv \ln \frac{-Q^2}{\mu^2} \,,
    &
    \Lx &\equiv \ln x \,,
    &
    \Lxb &\equiv \ln \bar{x} \,.  
\end{align}
We find
\begin{align} \label{eq:CA12}
    C_1^{(A1)} &= C_1^{(A1,A1)} = - C^{(A0)} \,,  
    \nonumber\\
    C_2^{(A1,A1)} &= 2 \, \frac{d}{dL} \spac C^{(A0)} = \frac{\alpha_s}{4\pi} \, C_F \bigg[ \frac{4}{\epsilon} - 4 \spac \lnQ + 6 \bigg] + \mathcal{O}(\alpha_s^2) \,,  
    \nonumber\\
    C_2^{(A1)} &= C_3^{(A1,A1)} = C_4^{(A1,A1)} = C_1^{(A2)} = C_2^{(A2)} = \mathcal{O}(\alpha_s^2) \,.  
\end{align}
Our results agree with known results in~ref.~\cite{Strohm:2020}. These results can be understood from reparameterization invariance (RPI)~\cite{Manohar:2002fd}, which relates short-distance coefficients of operators with different power counting. In particular, all $(An)$-type coefficients are determined in terms of the $(A0)$ one.

If one chooses to eliminate the operator $J_1^{(A1,A1)}$ with three $\gamma$ matrices  in terms of the evanescent operator $\hat{J}_1^{(A1,A1)}$ in~eq.~\eqref{eq:evanescent_operators}, the short-distance coefficients of the remaining operators get modified according to
\begin{align}
    C_2^{(A1,A1)} &\to C_2^{(A1,A1)} - C_1^{(A1,A1)} \,,
    &
    C_{3,4}^{(A1,A1)} &\to C_{3,4}^{(A1,A1)} + C_1^{(A1,A1)} \,.  
\end{align}

\subsubsection*{\boldmath Coefficients of $B$-type operators}

Projecting the right-hand side of~eq.~\eqref{eq:match-rel} with an external gluon-quark-antiquark state instead, one finds
\begin{align} \label{eq:B-RHS}
    &\int\!dt \, d^2s \, \sum_i C_i(t,s_1,s_2) \, \langle g(p_1) \bar{q}(p_2) q(k)| \spac J_i(t,s_1,s_2) \spac |0 \rangle
    \nonumber\\*
    &= - C_1^{(B1)}(\nm k,\np p_1,\np p_2) \, g_s \, \bar{u}_\Cb(k) \, \frac{\np^\mu}{\np(p_1+p_2)} \, \slashed{\varepsilon}_{\C\perp}^\ast(p_1) \, t^a \, v_\C(p_2) + \dots
    \nonumber\\*
    &\phantom{=}  - C_1^{(A1,B1)}(\nm k,\np p_1,\np p_2) \, g_s \, \bar{u}_\Cb(k) \, \frac{\slashed{k}_\perp \spac \gamma_\perp^\mu}{\nm k \, \np(p_1+p_2)} \, \slashed{\varepsilon}_{\C\perp}^\ast(p_1) \, t^a \, v_\C(p_2) + \dots
    \nonumber\\*
    &\phantom{=} - C_1^{(B2)}(\nm k,\np p_1,\np p_2) \, g_s \, \bar{u}_\Cb(k) \, \frac{\gamma_\perp^\mu \spac (\slashed{p}\vphantom{p}_{1\perp}+\slashed{p}\vphantom{p}_{2\perp})}{\nm k \, \np(p_1+p_2)} \, \slashed{\varepsilon}_{\C\perp}^\ast(p_1) \, t^a \, v_\C(p_2) + \dots \,,  
\end{align}
where $a$ is the colour of the gluon and quark colours are again kept implicit.
We show a representative contribution from $(B1)$, $(A1,B1)$, and $(B2)$-type operators.
The remaining terms, indicated by the ellipsis, can be obtained from~eqs.~\eqref{eq:J-B1}, \eqref{eq:J-A1B1} and~\eqref{eq:J-B2} by replacing derivatives similar to above.
In principle, also $\np\varepsilon_\C(p_1)$ terms contribute to the gluon-quark-antiquark matrix element.
We argue below why these can be ignored.
The momentum-space short-distance coefficients in~eq.~\eqref{eq:B-RHS} are defined by
\begin{align} \label{eq:CB-mom}
    C_i(\nm k,\np p_1,\np p_2) = \int\!dt \, d^2s \, C_i(t,s_1,s_2) \, e^{i \spac t \spac \nm k+i \spac s_1 \spac \np p_1 + i \spac s_2 \spac \np p_2} \,.  
\end{align}
It is convenient to introduce the momentum fraction $x\in[0,1]$ such that $\np p_1=x \, \np p$ and $\np p_2 = \bar{x} \, \np p$ where $\bar{x}\equiv 1-x$ and $p=p_1 + p_2$ is the total collinear momentum.\footnote{Since $\np p_i$ is the large collinear momentum component, the momentum fraction $x$ is of $\mathcal{O}(1)$ in the power expansion.} 
The short-distance coefficients can then be considered as functions of $x$ and $Q^2=\nm k \, \np p$.

\begin{figure}[t]
    \centering
    \begin{tabular}{ccc}
        \includegraphics[scale=0.48]{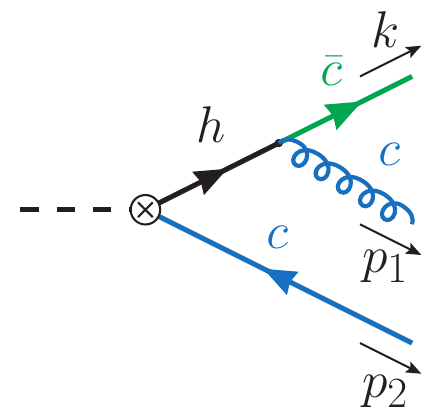} &
        \includegraphics[scale=0.48]{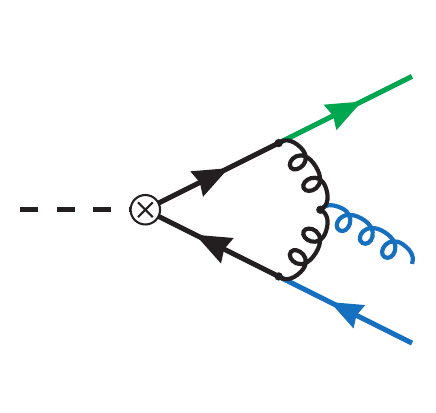} &
        \includegraphics[scale=0.48]{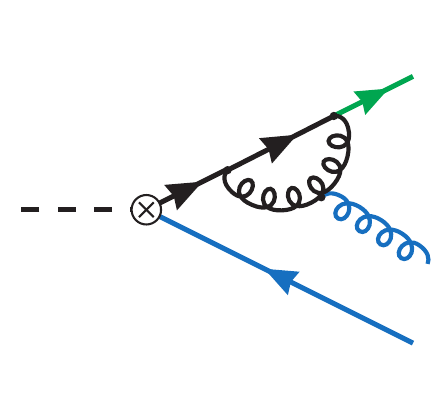} \\[-0.2cm]
        \includegraphics[scale=0.48]{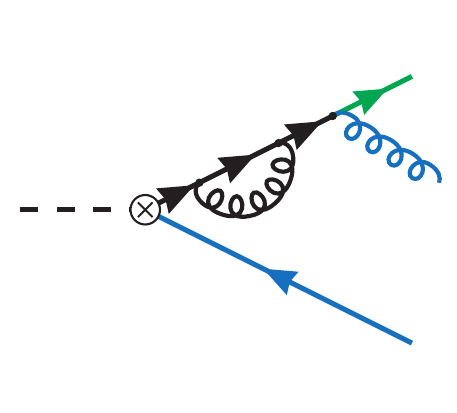} &
        \includegraphics[scale=0.48]{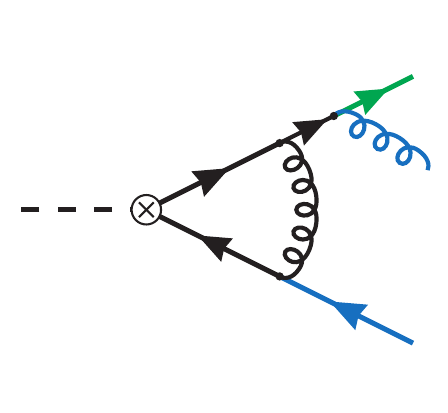} &
        \includegraphics[scale=0.48]{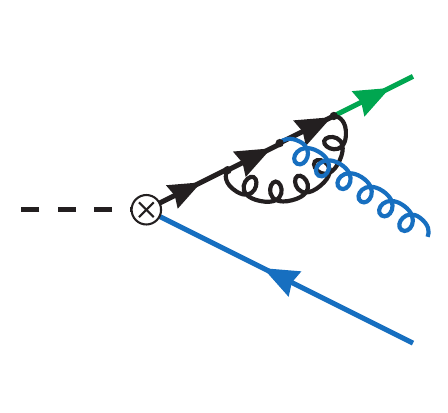} \\[-0.2cm]
        \includegraphics[scale=0.48]{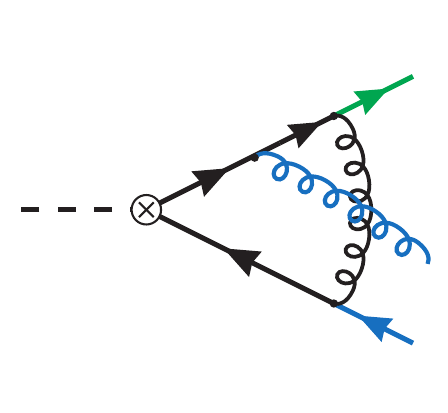} &
        \includegraphics[scale=0.48]{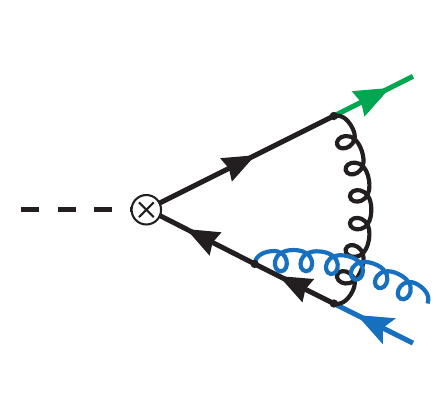} & 
    \end{tabular}
    \caption{Tree and one-loop diagrams contributing to the left-hand side of~eq.~\eqref{eq:match-rel} when projected with the matrix element $\langle g(p_1) \bar{q}(p_2) q(k)| \dots |0 \rangle$. The colours are the same as in fig.~\ref{fig:A-diagrams}. We do not show one-loop diagrams without hard region.}
    \label{fig:B-diagrams}
\end{figure}

When calculating the diagrams contributing to the matrix element on the left-hand side of~eq.~\eqref{eq:B-RHS}, we again employ eq.~\eqref{eq:QCD-SCET-spinor} and use the on-shell conditions for the momenta to eliminate the small momentum components.
Similarly, the on-shell condition of the polarization vector $p_1\cdot\varepsilon(p_1)=0$ can be used to eliminate
\begin{align}
    \nm \varepsilon(p_1) = \frac{p_{1\perp}^2 \, \np\varepsilon(p_1)}{(\np p_1)^2} - \frac{2 \, p_{1\perp}\cdot\varepsilon_\perp(p_1)}{\np p_1} \,.  
\end{align}
However, the first term is a contribution from the collinear Wilson lines inside the operator definitions.
Gauge invariance of the matching equation allows us thus to replace the QCD polarization vector by
\begin{align}
    \varepsilon^\mu(p_1) \to - \frac{p_{1\perp}\cdot\varepsilon_{\C\perp}(p_1) \, \np^\mu}{\np p_1} + \varepsilon_{\C\perp}^\mu(p_1) \,,  
\end{align}
and to ignore $\np\varepsilon_\C(p_1)$ contributions to the right-hand side of the matching equation. 
It is now straightforward to calculate the diagrams in fig.~\ref{fig:B-diagrams} and by comparing to~eq.~\eqref{eq:B-RHS} determine the $B$-type short-distance coefficients.

\begin{figure}[t]
    \centering
    \begin{tabular}{ccc}
        \includegraphics[scale=0.48]{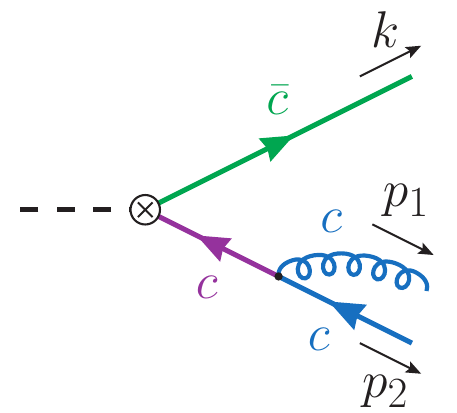} & \hspace*{12mm} & 
        \includegraphics[scale=0.48]{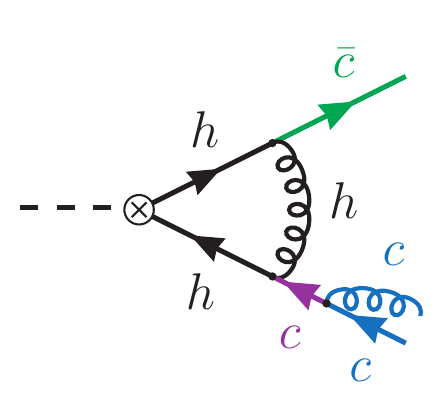} \\
        (a) && (b) 
    \end{tabular}
    \caption{Tree and one-loop diagrams contributing to the left-hand side of~eq.~\eqref{eq:match-rel} when projected with the matrix element $\langle g(p_1) \bar{q}(p_2) q(k)| \dots |0 \rangle$. The colours are the same as in fig.~\ref{fig:A-diagrams}. Even though the purple propagator is collinear these diagrams contribute to the matching coefficients as described in the main text.}
    \label{fig:B-diagrams-onshell}
\end{figure}

The two diagrams depicted in fig.~\ref{fig:B-diagrams-onshell} also contribute to the on-shell matching even though the purple propagator is collinear and not hard.
To understand this, we study them in more detail by writing 
\begin{align} \label{eq:B-onshell-diag}
    \langle g(p_1) \bar{q}(p_2) q(k)| \spac \bar\psi \spac \gamma^\mu \spac \psi(0) \spac |0 \rangle \big|_\mathrm{fig.\spac\ref{fig:B-diagrams-onshell}} = ig_s \, \bar{u}(k) \, \Lambda^\mu (p^2, p\!\cdot\!k) \, \frac{i\slashed{p}}{p^2} \, \slashed{\varepsilon}^*(p_1) \, t^a \, v(p_2) \,,  
\end{align}
where $p^2=(p_1+p_2)^2\neq0$ and $\Lambda^\mu$ represents the one-particle-irreducible (1PI) part of the diagrams.
In~ref.~\cite{Beneke:2004rc}, a method was proposed to separate the ``local'' terms, which contribute to the short-distance coefficients at $\mathcal{O}(\lambda)$, from the non-local terms, which are reproduced by time-ordered products of $(An)$-type operators with leading-power SCET Lagrangian insertions.
The trick is to decompose the propagator as
\begin{align} \label{eq:propagator_decomp}
    \frac{i\slashed{p}}{p^2} = \frac{i}{\np p} \, \frac{\nps}{2} + \frac{i}{p^2} \bigg(\np p \, \frac{\nms}{2} - \frac{p_\perp^2}{\np p} \, \frac{\nps}{2} + \slashed{p}\vphantom{p}_\perp \bigg) \,.  
\end{align}
When inserted into eq.~\eqref{eq:B-onshell-diag}, the second term on the right-hand side yields non-local contributions up to $\mathcal{O}(\lambda)$ in the power expansion.
To this order, it is thus sufficient to consider the first term only.
However, expanding to $\mathcal{O}(\lambda^2)$ generates terms proportional to $p^2$ which cancel the denominator of the second term in~eq.~\eqref{eq:propagator_decomp} and render it local.
To obtain the expansion of $\Lambda^\mu$ one needs to evaluate the two diagrams in fig.~\ref{fig:A-diagrams} for an \emph{off-shell} collinear momentum $p^2\neq0$, see appendix~\ref{app:MoR-trick} for details.
Instead of considering the diagrams in fig.~\ref{fig:B-diagrams-onshell}, one can apply the equations of motion on operator level, i.e.\ perform off-shell matching.
This is discussed in appendix~\ref{app:off-shell-match}.

Proceeding as explained, we find for matching coefficients of the $\mathcal{O}(\lambda)$ operators~\eqref{eq:J-B1}
\begin{align} \label{eq:CB1-bare}
    -C_1^{(B1)} = C_2^{(B1)} &= 1 + \frac{\alpha_s}{4\pi} \, C_A  \bigg[ - \frac{1}{\epsilon} \frac{\Lxb}{x} + \frac{(\lnQ-2) \spac \Lxb}{x} +\frac{\Lsqxb}{2 x} - \frac{\Lx}{\bar{x}} \bigg]  
    \nonumber\\
    &\hspace{7.5mm} + \frac{\alpha_s}{4\pi} \, C_F \bigg[ -\frac{2}{\epsilon^2} + \frac{2 \spac \lnQ - 1}{\epsilon} + \frac{2}{\epsilon} \frac{\Lxb}{x} - \frac{2(\lnQ - 2) \spac \Lxb}{x} - \frac{\Lsqxb}{x} 
    \nonumber\\*
    &\hspace{26mm} - \lnsqQ + \lnQ - 3 + \frac{\pi^2}{6} \bigg] + \mathcal{O}(\alpha_s^2) \,,  
\end{align}
where $\lnQ, \Lx$ and $\Lxb$ are defined in~eq.~\eqref{eq:logs}.
Our result agrees with the one found in refs.~\cite{Strohm:2020,Vladimirov:2021hdn}.

At order $\mathcal{O} (\lambda^2)$, it turns out that the short-distance coefficients of the four $(A1,B1)$-type operators~\eqref{eq:J-A1B1} can be expressed as
\begin{align} \label{eq:CA1B1}
    C^{(A1,B1)}_1 &= - C^{(A0)} \,,
    &
    C^{(A1,B1)}_2 &= \mathcal{O}(\alpha_s^2) \,,  
    \nonumber\\
    C^{(A1,B1)}_3 &= 2 \big[ C^{(A0)} + C_1^{(B1)} \big] \,, 
    &
    C^{(A1,B1)}_4 &= \frac{2}{x} \, C^{(A0)} \,,  
\end{align}
with the two relevant coefficients given in~eqs.~\eqref{eq:CA0} and~\eqref{eq:CB1-bare}.

For the eight $(B2)$-type operators~\eqref{eq:J-B2}, we find that only half of the short-distance coefficients are independent, namely 
\begin{align} \label{eq:CB2-1-to-4}
    C_1^{(B2)} &= -\frac{1}{x} + \frac{\alpha_s}{4\pi} \, C_A
    \bigg[ \frac{1}{\epsilon} \bigg(\frac{\Lxb}{x} + \frac{3\spac \Lx}{\bar{x}} \bigg) - \frac{3(\lnQ-2) \spac \Lx}{\bar{x}} - \frac{(\lnQ-6) \spac \Lxb}{x} - \frac{3\spac \Lsqx}{2\bar{x}} - \frac{\Lsqxb}{2 x} - \frac{1}{x} \bigg]  
    \nonumber\\
    &\quad + \frac{\alpha_s}{4\pi} \, C_F \bigg[ \frac{1}{\epsilon^2} \frac{2}{x}  - \frac{1}{\epsilon} \bigg(\frac{2 \spac \lnQ  + 2\spac \Lxb -3}{x} - 3 \bigg) + \frac{2(\lnQ-6)\spac \Lxb}{x} + \frac{3\spac \Lx}{\bar{x}} + \frac{\Lsqxb}{x}
    \nonumber\\*
    &\hspace{22mm} + \frac{1}{x} \bigg( \lnsqQ -3\spac \lnQ + 9 - \frac{\pi^2}{6} \bigg) - 3(\lnQ-1) \bigg] + \mathcal{O}(\alpha_s^2) \,,  
    \nonumber\\[4mm]
    C_2^{(B2)} &= -\frac{2}{x} - \frac{\alpha_s}{4\pi} \, C_A \bigg[\frac{1}{\epsilon} \bigg(\frac{(1 - 3 x)\Lxb}{x^2} - \frac{3 \Lx}{\bar{x}} + \frac{1}{x} \bigg) + \lnQ \bigg(\frac{3 \Lx}{\bar{x}} - \frac{(1 - 3 x) \Lxb}{x^2} - \frac{1}{x} \bigg) 
    \nonumber\\*
    &\hspace{30mm} - \frac{(1 - 3 x) \Lsqxb}{2 x^2} + \frac{(5 - 17 x) \Lxb}{2 x^2} + \frac{3 \Lsqx}{2 \bar{x}} - \frac{15 \Lx}{2\bar{x}} + \frac{9}{2 x} \bigg]  
    \nonumber\\
    &\quad + \frac{\alpha_s}{4\pi} \, C_F \bigg[ \frac{1}{\epsilon^2} \frac{4}{x} - \frac{1}{\epsilon} \bigg(\frac{4 \spac \lnQ -8}{x} - \frac{2 (1 - 3 x)\Lxb}{x^2} \bigg) + \frac{2 \spac \lnsqQ}{x} - \lnQ \bigg(\frac{2 (1 - 3 x) \Lxb}{x^2} + \frac{8}{x} \bigg)
    \nonumber\\*
    &\hspace{24mm}  - \frac{(1 - 3 x) \Lsqxb}{x^2} + \frac{(5 - 17 x) \Lxb}{x^2} + \frac{5 \Lx}{\bar{x}} + \frac{23}{x} - \frac{\pi^2}{3x} \bigg] + \mathcal{O}(\alpha_s^2) \,,  
    \nonumber\\[4mm]
    C_3^{(B2)} &= \frac{2}{x} - \frac{\alpha_s}{4\pi} \, C_A \bigg[\frac{1}{\epsilon} \bigg(\frac{(1 + x) \Lxb}{x^2} + \frac{3 \Lx}{\bar{x}} + \frac{1}{x}\bigg) - \lnQ \bigg(\frac{(1+x) \Lxb}{x^2} + \frac{3 \Lx}{\bar{x}} + \frac{1}{x}\bigg)
    \nonumber\\*
    &\hspace{27mm} - \frac{(1 + x) \Lsqxb}{2 x^2} + \frac{(5 + 7 x) \Lxb}{2 x^2} - \frac{3 \Lsqx}{2 \bar{x}} +\frac{13 \Lx}{2 \bar{x}} + \frac{1}{2 x} \bigg]  
    \nonumber\\
    &\quad - \frac{\alpha_s}{4\pi} \, C_F \bigg[ \frac{1}{\epsilon^2} \frac{4}{x} + \frac{1}{\epsilon} \bigg(- \frac{4 \spac \lnQ - 4}{x} - \frac{2 (1 + x)\Lxb}{x^2} + 2 \bigg) + \frac{2 \spac \lnsqQ}{x}
    \nonumber\\*
    &\hspace{22mm} + \lnQ \bigg(\frac{2 (1 + x) \Lxb}{x^2}-\frac{2 (2 + x)}{x} \bigg) + \frac{(1+x) \Lsqxb}{x^2} 
    \nonumber\\*
    &\hspace{22mm} -\frac{(5 + 7 x) \Lxb}{x^2} + \frac{5 \Lx}{\bar{x}} - \frac{\pi^2}{3x} + \frac{13}{x} + 6 \bigg] + \mathcal{O}(\alpha_s^2) \,,  
    \nonumber\\[4mm]
    C_4^{(B2)} &= \frac{2}{x} + \frac{\alpha_s}{4\pi} \, C_A \bigg[ \frac{1}{\epsilon} \bigg(\frac{(1 - 3 x) \Lxb}{x^2} - \frac{3 \Lx}{\bar{x}} +\frac{1}{x} \bigg) - \lnQ \bigg(\frac{(1 - 3 x) \Lxb}{x^2} - \frac{3 \Lx}{\bar{x}} +\frac{1}{x} \bigg)
    \nonumber\\*
    &\hspace{29mm} - \frac{(1 - 3 x) \Lsqxb}{2 x^2} + \frac{3 \Lsqx}{2 \bar{x}} - \frac{5 (1 + 3 x) \Lxb}{2 x^2} - \frac{13 \Lx}{2 \bar{x}} + \frac{7}{2 x} \bigg]  
    \nonumber\\
    &\hspace{4mm} - \frac{\alpha_s}{4\pi} \, C_F  \bigg[ \frac{1}{\epsilon^2} \frac{4}{x} + \frac{1}{\epsilon} \bigg(-\frac{4 \spac \lnQ - 8}{x} + \frac{2 (1 - 3 x) \Lxb}{x^2}\bigg) +\frac{2 \spac \lnsqQ}{x} - \lnQ \bigg(\frac{2 (1 - 3 x) \Lxb}{x^2} + \frac{8}{x}\bigg)
    \nonumber\\*
    &\hspace{23mm} - \frac{(1 - 3 x) \Lsqxb}{x^2} - \frac{5 (1 + 3 x) \Lxb}{x^2} - \frac{(4 - 5 x) \Lx}{x \bar{x}} - \frac{\pi ^2}{3 x} + \frac{11}{x} \bigg] + \mathcal{O}(\alpha_s^2) \,.  
\end{align}
The matching coefficients of the remaining $(B2)$-type operators can be expressed in terms of these coefficients and the one given in~eq.~\eqref{eq:CB1-bare}.
The relations read
\begin{align} \label{eq:CB2-5-to-8}
    C^{(B2)}_5 &= \frac{1}{\bar{x}} \, C_1^{(B2)} \,,
    &
    C^{(B2)}_6 &= - \frac{1}{\bar{x}} \, C_2^{(B2)} \,,  
    \nonumber\\
    C^{(B2)}_7 &= -\frac{1}{\bar{x}} \big[ 2 \spac C_1^{(B1)} + C_3^{(B2)} \big] \,,
    &
    C^{(B2)}_{8} &= - \frac{1}{\bar{x}} \big[ 2 \spac C_1^{(B2)} + C_4^{(B2)} \big] \,.  
\end{align}
As was the case for eq.~\eqref{eq:CA12}, the relations for the $B$-type coefficients presented in eqs.~\eqref{eq:CA1B1} and~\eqref{eq:CB2-5-to-8} are in fact a consequence of RPI.\footnote{We checked that  eqs.~\eqref{eq:CA12},~\eqref{eq:CA1B1},~\eqref{eq:CB2-5-to-8} satisfy the RPI relations derived by J.~Strohm (unpublished notes, 2023).}

If one eliminates the operators $J_1^{(A1,B1)}$, $J_1^{(B2)}$, $J_5^{(B2)}$ containing three $\gamma$ matrices in favour of the evanescent operators defined in~eq.~\eqref{eq:evanescent_operators}, the short-distance coefficients of the remaining operators are modified according to
\begin{align}
    C_{2,3}^{(A1,B1)} &\to C_{2,3}^{(A1,B1)} + C_1^{(A1,B1)} \,,
    &
    C_4^{(A1,B1)} &\to C_4^{(A1,B1)} - C_1^{(A1,B1)} \,,  
    \nonumber\\
    C_2^{(B2)} &\to C_2^{(B2)} - C_1^{(B2)} \,,
    &
    C_{3,4}^{(B2)} &\to C_{3,4}^{(B2)} + C_1^{(B2)} \,,  
    \nonumber\\
    C_7^{(B2)} &\to C_7^{(B2)} - C_5^{(B2)} \,,
    &
    C_{6,8}^{(B2)} &\to C_{6,8}^{(B2)} + C_5^{(B2)} \,.  
\end{align}

\section{Renormalization}
\label{sec:renormalization}

The renormalization of the matching coefficients is most conveniently discussed in momentum space.
The defining relation~\eqref{eq:match-rel} reads in this case
\begin{align} \label{eq:match-rel-mom}
    \bar\psi \spac \gamma^\mu \spac \psi(0) &= \int\!\frac{d\nm k}{2\pi} \, \frac{d\np p}{2\pi} \, \sum_i C_i(\nm k,\np p) \, J_i(\nm k,\np p)
    \nonumber\\*
    & + \int\!\frac{d\nm k}{2\pi} \, \frac{d\np p}{2\pi} \, \frac{\np p}{2\pi}  \int_0^1\!dx \sum_i C_i(\nm k,\np p,x) \, J_i(\nm k,\np p,x) + \dots \,, 
\end{align}
where we dropped terms irrelevant for this work and the factor $\np p$ is the Jacobian from the variable change $(\np p_1,\np p_2)\to(\np p,x)$.
The momentum-space operators are defined similarly to~eqs.~\eqref{eq:CA-mom} and~\eqref{eq:CB-mom}.
As the electro-magnetic current is conserved, eq.~\eqref{eq:match-rel-mom} also holds for renormalized quantities with
\begin{align} \label{eq:ren-J-C}
    J_j^\mathrm{ren}(\{\underline{x}\}) &= \sum_k \int\!d{\{\underline{y}\}} \, Z_{jk}(\{\underline{x}\},\{\underline{y}\}) \, J_k(\{\underline{y}\}) \,,  
    \nonumber\\*
    C_j^\mathrm{ren}(\{\underline{x}\}) &= \sum_i \int\!d{\{\underline{y}\}} \,  C_i(\{\underline{y}\}) \, Z_{ij}^{-1}(\{\underline{y}\},\{\underline{x}\}) \,,  
\end{align}
where $\{\underline{x}\},\,\{\underline{y}\}$ denote sets of momentum fractions, which are empty for $A$-type operators.
Here and in the following, we drop the $\nm k$, $\np p$ arguments to increase readability.
The $Z$ factor up to $\mathcal{O}(\lambda^2)$ was determined in~refs.~\cite{Beneke:2017ztn,Beneke:2018rbh}.
Operators of the same order in $\lambda$ mix under renormalization and therefore also their short-distance coefficients.

\subsubsection*{\boldmath $\mathcal{O}(\lambda^0)$ renormalization}
At leading power, only the single operator~\eqref{eq:J-A0} contributes.
The renormalization in this case is multiplicative with~\cite{Bauer:2003di}
\begin{align} \label{eq:Z-A0}
    Z^{(A0),(A0)} = 1 - \frac{\alpha_s}{4\pi} \, C_F \bigg[ \frac{2}{\epsilon^2} - \frac{2}{\epsilon} \, \lnQ + \frac{3}{\epsilon} \bigg] + \mathcal{O}(\alpha_s^2) \,.  
\end{align}
It is easy to check that the inverse of this $Z$ factor removes the singular parts in~eq.~\eqref{eq:CA0}.

\subsubsection*{\boldmath $\mathcal{O}(\lambda)$ renormalization}

As there are four operators at $\mathcal{O}(\lambda)$ in the power expansion, the $Z$ factor is a $4\times4$ matrix.
Its general structure is~\cite{Beneke:2017ztn,Beneke:2018rbh}
\begin{align}
    \left(
    \begin{array}{cc}
        Z^{(A1),(A1)} & 0_{2\times2} \\
        0_{2\times2} & Z^{(B1),(B1)}
    \end{array} \right) \,,  
\end{align}
where each entry is a $2\times2$ matrix.
The renormalization of the $(A1)$-type operators~\eqref{eq:J-A1} is again multiplicative with $Z^{(A1),(A1)}=Z^{(A0),(A0)} \, \mathbbm{1}_{2\times2}$, i.e.\ the same as at leading power.
For the $(B1)$-type operators~\eqref{eq:J-B1}, the renormalization condition~\eqref{eq:ren-J-C} contains a convolution in the momentum fraction $y$ 
with
\begin{align} \label{eq:Z-B1}
    &Z_{ij}^{(B1),(B1)}(x,y) 
    \nonumber\\*
    &= \delta(x-y) \spac \delta_{ij} - \frac{\alpha_s}{4\pi} \,\Bigg\{ C_F \bigg[ \frac{2}{\epsilon^2} - \frac{2}{\epsilon} \, \ln\frac{\bar{x} Q^2}{-\mu^2} + \frac{3}{\epsilon} \bigg] + \frac{C_A}{\epsilon} \, \ln\frac{\bar{x}}{x} \Bigg\} \, \delta(x-y) \spac \delta_{ij}
    \nonumber\\*[1mm]
    &\hspace{25mm} + \frac{1}{\epsilon} \, \gamma^{\A\Ch,\A\Ch}(x,y) \spac \delta_{ij} + \mathcal{O}(\alpha_s^2) \,,  
\end{align}
where $i,j=1,2$ and 
\begin{align} \label{eq:gamma-ACh-ACh}
    \gamma^{\A\Ch,\A\Ch}(x,y) ={}& -\frac{\alpha_s}{4\pi} \, C_A\, \Bigg\{ \bigg[\frac{\theta(x-y)}{x-y}\bigg]_+ + \bigg[\frac{\theta(y-x)}{y-x}\bigg]_+ - \frac{\theta(x-y)}{x\bar{y}} \, (x^2 + \bar{x} \bar{y})
    \nonumber\\*
    &\hspace{18mm} - \frac{\theta(y-x)}{y} \, (\bar{x}-y) \Bigg\}
    \nonumber\\*
    & - \frac{\alpha_s}{2\pi} \,\bigg(C_F-\frac{C_A}{2}\bigg) \,\Bigg\{ \frac{\theta(x-\bar{y}) \spac \bar{x}}{xy} + \frac{\theta(\bar{y}-x)}{\bar{y}} \Bigg\} \,(\bar{x}-y) + \frac{\alpha_s}{2\pi} \, C_F \, \bar{x} \,.  
\end{align}
The plus distribution is defined as
\begin{align}
    \int_0^1\!dy \, \big[\mathcal{K}(x,y)\big]_+ \, f(y) &= \int_0^1\!dy \, \mathcal{K}(x,y) \, \big[f(y)-f(x)\big] \,.  
\end{align}
To obtain eq.~\eqref{eq:gamma-ACh-ACh}, one has to project equation~(C.2) in~ref.~\cite{Beneke:2017ztn} with the proper Dirac structures.
It is crucial to project in $d$ dimensions, then drop finite terms of $\mathcal{O}(\epsilon^0)$ in accordance with $\overline{\text{MS}}$ renormalization.
Convoluting~eq.~\eqref{eq:CB1-bare} with the inverse of~eq.~\eqref{eq:Z-B1} in $y$ removes all singular terms.

\subsubsection*{\boldmath $\mathcal{O}(\lambda^2)$ renormalization}

The relevant part of the $Z$ factor at $\mathcal{O}(\lambda^2)$ has the general form~\cite{Beneke:2017ztn,Beneke:2018rbh}
\begin{align} \label{eq:Z-NNLP}
    \left(
    \begin{array}{ccc|ccc}
        Z^{(A2),(A2)} & 0 & 0 & 0 & 0 & 0 \\
        0 & Z^{(B2),(B2)} & Z^{(B2),(C2)}\, & 0 & 0 & 0 \\
        0 & 0 & Z^{(C2),(C2)} \;& 0 & 0 & 0 \\[0.1cm] \hline
        &&&&&\\[-0.4cm]
        0 & 0 & 0 & \,Z^{(A1,A1),(A1,A1)} & 0 & 0 \\
        0 & 0 & 0 & 0 & Z^{(A1,B1),(A1,B1)} & 0 \\
        0 & 0 & 0 & \,Z^{(T1,T1),(A1,A1)} & Z^{(T1,T1),(A1,B1)} & Z^{(T1,T1),(T1,T1)}
    \end{array} \right) \,.  
\end{align}
When renormalizing the operators defined in~eqs.~\eqref{eq:J-A1A1} to~\eqref{eq:J-B2} one also needs to include the two time-ordered product operators
\begin{align} \label{eq:J-T1T1}
    J_1^{(T1,T1)}(t,s) &= i^2\!\int\!d^dz_1 \int\!d^dz_2 \, \text{T}\big\{\bar\Ch_\Cb(t\nm),\mathcal{L}_{\bar\xi}^{(1)}(z_1) \big\} \, \gamma_\perp^\mu \, \text{T}\big\{\Ch_\C(s\np), \mathcal{L}_\xi^{(1)}(z_2)\big\} \,,  
    \nonumber\\*
    J_2^{(T1,T1)}(t,s) &= i^2\!\int\!d^dz_1 \int\!d^dz_2 \, \text{T}\big\{\bar\Ch_\Cb(t\nm),\mathcal{L}_{\bar\xi}^{(1)}(z_1) \big\} \, \gamma_\perp^\mu \, \text{T}\big\{\Ch_\C(s\np), \mathcal{L}_\mathrm{YM}^{(1)}(z_2)\big\} \,,  
\end{align}
where $\mathcal{L}_\xi^{(1)}$ and $\mathcal{L}_\mathrm{YM}^{(1)}$ are subleading-power SCET Lagrangians~\cite{Beneke:2002ni}.
Their short-distance coefficients are given by $C_1^{(T1,T1)}=C_2^{(T1,T1)}=C^{(A0)}$ to all orders in perturbation theory as the SCET Lagrangian does not renormalize~\cite{Beneke:2002ph}.
In principle, one also needs to include $(A0,C2)$-type operators \cite{Beneke:2018rbh} . 
However, when focusing on the short-distance coefficients, they can be ignored due to the upper triangular structure in~eq.~\eqref{eq:Z-NNLP}.
Operators of $(B1,B1)$ type mix with the time-ordered product operators~\eqref{eq:J-T1T1} but are irrelevant for the renormalization of the short-distance coefficients considered in this work.

The renormalization of the $A$-type operators is multiplicative and again given by the leading power $Z$ factor~\eqref{eq:Z-A0}, i.e.\ one has
\begin{align} \label{eq:Z-A2}
    Z^{(A2),(A2)} &= Z^{(A0),(A0)} \, \mathbbm{1}_{2\times2} \,,
    &
    Z^{(A1,A1),(A1,A1)} &= Z^{(A0),(A0)} \, \mathbbm{1}_{4\times4} \,.  
\end{align}
The $(A1,A1)$-type operators also mix with the time-ordered product operators~\eqref{eq:J-T1T1}.
This mixing is described by~\cite{Beneke:2018rbh}
\begin{align}
    Z_{kj}^{(T1,T1),(A1,A1)} = \frac{\alpha_s}{\pi} \, \frac{C_F}{\epsilon} \, \delta_{k1} \spac \delta_{j2} + \mathcal{O}(\alpha_s^2) \,,  
\end{align}
with $j=1,2,3,4$ and $k=1,2$.
The renormalized short-distance coefficients are then given by
\begin{align}
    C^{(A1,A1)}_{\mathrm{ren},j} = \sum_{i=1}^4 C_i^{(A1,A1)} \, (Z_{ij}^{-1})^{(A1,A1),(A1,A1)} + \sum_{k=1}^2 C^{(A0)} \, (Z_{kj}^{-1})^{(T1,T1),(A1,A1)} \,.  
\end{align}
Plugging in the results~\eqref{eq:CA0} and~\eqref{eq:CA12} for the bare short-distance coefficients, all divergent parts are removed.

For the $(B2)$-type operators, one finds
\begin{align} \label{eq:Z-B2}
    Z_{ij}^{(B2),(B2)}(x,y) 
    = \delta(x-y) \spac \delta_{ij} &- \frac{\alpha_s}{4\pi} \,\Bigg\{ C_F \bigg[ \frac{2}{\epsilon^2} - \frac{2}{\epsilon} \, \ln\frac{\bar{x}Q^2}{-\mu^2} + \frac{3}{\epsilon} \bigg] + \frac{C_A}{\epsilon} \, \ln\frac{\bar{x}}{x} \Bigg\} \, \delta(x-y) \spac \delta_{ij}
    \nonumber\\*[1mm]
    &+ \frac{1}{\epsilon} \, \gamma_{ij}^{(B2),(B2)}(x,y) + \mathcal{O}(\alpha_s^2) \,,  
\end{align}
where the only difference to~eq.~\eqref{eq:Z-B1} is that the anomalous dimension $\gamma^{(B2),(B2)}$ is an $8\times8$ matrix.
It is convenient to distinguish two kinds of operators:
the ones where the derivative acts on the collinear quark field only and those where it acts on both fields.
The anomalous dimension then takes the form
\begin{align} \label{eq:gamma_B2_split}
    \gamma^{(B2),(B2)} = \left(
    \begin{array}{cc}
        \gamma^{\partial[\A\Ch],\partial[\A\Ch]} & 0_{4\times4} \\
        \gamma^{\A\partial\Ch,\partial[\A\Ch]} & \gamma^{\A\partial\Ch,\A\partial\Ch}
    \end{array}
    \right) \,,  
\end{align}
where each entry is a $4\times4$ matrix and given as function of the momentum fractions $x,y$ in appendix~\ref{app:anomalous-dim}.  
The short-distance coefficients are then renormalized as
\begin{align}
    C^{(B2)}_{\mathrm{ren},\spac j}(x) = \sum_{i=1}^8 \int_0^1\!dy \, C^{(B2)}_i(y) \, (Z^{-1})_{ij}^{(B2),(B2)}(y,x) \,, \qquad j=1,2,3,4 \,,  
    \nonumber\\
    C^{(B2)}_{\mathrm{ren},\spac j}(x) = \sum_{i=5}^8 \int_0^1\!dy \, C^{(B2)}_i(y) \, (Z^{-1})_{ij}^{(B2),(B2)}(y,x) \,, \qquad j=5,6,7,8 \,.  
\end{align}
Again plugging in the results for the bare short-distance coefficient in eqs.~\eqref{eq:CB2-1-to-4} and~\eqref{eq:CB2-5-to-8} and performing the convolution, all divergences are cancelled.\footnote{Even though individual terms in $\gamma^{(B2),(B2)}$ give rise to divergent convolutions, in sum the result is well defined.}

For the $(A1,B1)$-type operators, there are two relevant $Z$ factors.
The first one is
\begin{align} \label{eq:Z-A1B1}
    &Z_{ij}^{(A1,B1),(A1,B1)}(x,y) 
    \nonumber\\*
    ={}& \delta(x-y) \spac \delta_{ij} - \frac{\alpha_s}{4\pi} \,\Bigg\{ C_F \bigg[ \frac{2}{\epsilon^2} - \frac{2}{\epsilon} \, \ln\frac{\bar{x}Q^2}{-\mu^2} + \frac{3}{\epsilon} \bigg] + \frac{C_A}{\epsilon} \, \ln\frac{\bar{x}}{x} \Bigg\} \, \delta(x-y) \spac \delta_{ij}
    \nonumber\\*
    &\hspace{20mm} + \frac{1}{\epsilon} \, \gamma_{ij}^{\A\Ch,\A\Ch}(x,y) + \mathcal{O}(\alpha_s^2) \,,  
\end{align}
with $i,j=1,2,3,4$.
The term in the second line is again related to equation~(C.2) in~ref.~\cite{Beneke:2017ztn}.
Its entries as function of the momentum fractions $x,y$ can also be found in appendix~\ref{app:anomalous-dim}.
The second $Z$ factor is~\cite{Beneke:2018rbh}
\begin{align}
    Z_{1j}^{(T1,T1),(A1,B1)}(y) &= + \frac{\alpha_s}{\pi} \, \frac{1}{2\epsilon} \bigg(C_F-\frac{C_A}{2}\bigg) \bigg[ \frac{y}{\bar{y}} \, \delta_{j1} - \frac{2y}{\bar{y}} \,  \delta_{j3} + \frac{2}{\bar{y}} \, \delta_{j4} \bigg] + \mathcal{O}(\alpha_s^2) \,,  
    \nonumber\\*[2mm]
    Z_{2j}^{(T1,T1),(A1,B1)}(y) &= - \frac{\alpha_s}{\pi} \, \frac{1}{2\epsilon} \, \frac{C_A}{2} \bigg[ \delta_{j1} - 2 \spac \delta_{j3} + \frac{2}{y} \, \delta_{j4} \bigg] + \mathcal{O}(\alpha_s^2) \,,  
\end{align}
for $j=1,2,3,4$.
We find for the renormalized short-distance coefficients
\begin{align}
    C^{(A1,B1)}_{\mathrm{ren},j}(x) ={}& \sum_{i=1}^4 \int_0^1\!dy \, C_i^{(A1,B1)}(y) \, (Z_{ij}^{-1})^{(A1,B1),(A1,B1)}(y,x)
    \nonumber\\*
    +{}& \sum_{k=1}^2 C^{(A0)} \, (Z_{kj}^{-1})^{(T1,T1),(A1,B1)}(x) \,, \qquad j=1,2,3,4 \,.  
\end{align}
All divergences are removed by plugging in the results~\eqref{eq:CA0} and~\eqref{eq:CA1B1} for the bare short-distance coefficients and performing the convolution on the right-hand side.


\section{Refactorization}
\label{sec:refactorization}

It is well known that individual subleading-power contributions to QCD processes are fraught with endpoint-divergent convolution integrals  
\cite{Geshkenbein:1982zs,Beneke:2000ry,Beneke:2003pa,Beneke:2008pi,Beneke:2019oqx,Moult:2019uhz}.  
However, in sum these divergences have to cancel by consistency in complete factorization theorems.
This cancellation is ensured by so-called \emph{refactorization} conditions for short-distance coefficients and operator matrix elements~\cite{Beneke:2008pi,Beneke:2020ibj,Liu:2020wbn,Beneke:2022obx}.
In this section, we show that the $C_i^{(B2)}$ determined above indeed fulfil such conditions.

For this discussion, it is convenient to change the basis of the $(B2)$-type operators to the one given in appendix~\ref{app:bases-conv}, with operators denoted by $\mathcal{J}_i^{(B2)}$ and  short-distance coefficients by $\mathcal{C}_i^{(B2)}$.
A typical factorization theorem contains these operators at amplitude level in convolutions of the form
\begin{align} \label{eq:NLP-factorization}
    &\int\!dt\,d^2s \, \sum_{i=1}^8 \mathcal{C}_i^{(B2)}(t,s_1,s_2) \, \big\langle \mathcal{J}_i^{(B2)}(t,s_1,s_2) \big\rangle
    \nonumber\\*
    & + \int\!du\,ds \, C^{(A0)}(u,s) \int\!dt\,dv \, \sum_{k=1}^2 D_k^{(B2)}(u,t,v) \, \big\langle O_k^{(B2)}(t,v) \, \gamma_\perp^\mu \, \Ch_\C(s\np) \big\rangle\,,  
\end{align}
where $\langle\ldots\rangle$ denotes some matrix element.
Analogous to the collinear functions introduced in refs.~\cite{Beneke:2018gvs,Beneke:2019oqx}, the coefficients $D_k^{(B2)}$ are defined by the position-space matching equation
\begin{align}
    i \! \int \! d^dz \, \text{T}\big\{ \bar\Ch_\Cb (u \nm), \mathcal{L}_{\bar\xi}^{(2)}(z) \big\} = \int\!dt\,dv \, \sum_{k=1}^2 D_k^{(B2)}(u,t,v) \, O_k^{(B2)}(t,v) \,,  
    \end{align}
where $\mathcal{L}_{\bar\xi}^{(2)}$ is a subleading-power SCET Lagrangian~\cite{Beneke:2002ni} and the operators on the right-hand side are
\begin{align} \label{eq:DB2-oper}
    O_1^{(B2)}(t,v) &= \bar\Ch_\Cb (t\nm) \, [\gamma_\perp^\rho, \gamma_\perp^\sigma] \, i\partial_{\perp\rho} \, \mathcal{B}_{\perp\sigma}(v \np) \,,  
    \nonumber\\
    O_2^{(B2)}(t,v) &= \bar\Ch_\Cb (t\nm) \, i\partial_\perp^\rho \, \mathcal{B}_{\perp\rho}(v \np) \,  
\end{align}
up to operators that vanish by the equations of motion. 
The soft-gluon building block is defined as~\cite{Beneke:2019oqx}
\begin{align}
    \mathcal{B}^\mu &\equiv Y^\dagger g_s \spac A_s^\mu \, Y + Y^\dagger [i\partial^\mu, Y] \,,
    &
    Y(x) &= \text{P} \exp\!\bigg[i g_s \int_{-\infty}^{0}\!\!ds \, \np A_s(x + s \np) \bigg] \,,  
\end{align}
where the soft Wilson lines $Y$ ensure soft gauge invariance.
The coefficients $D_k^{(B2)}$ can be regarded as soft-collinear splitting amplitudes that describe the splitting of an energetic, slightly off-shell parton into an energetic and a soft parton carrying light-cone momentum~$\omega$.
Their Fourier transforms 
\begin{align}
    D_k^{(B2)}(\nm q,\nm k,\omega) = \int\!du\,dt\,dv \, D_k^{(B2)}(u,t,v) \, e^{-i\spac u \spac \nm q+i\spac t \spac \nm k+i\spac v \spac \omega}  
\end{align}
were calculated in momentum space to the one-loop order in~ref.~\cite{Beneke:2019oqx}.
We state the results here for completeness\footnote{In the notation of~ref.~\cite{Beneke:2019oqx} one has $\omega D_1^{(B2)}=2\pi\spac J_6$ and $\omega D_2^{(B2)}=2\pi\spac J_1$, as the (anti-)collinear momenta are outgoing instead of incoming.}
\begin{align} \label{eq:DB2}
    D_1^{(B2)} &= - \frac{1}{2} \, \frac{1}{\nm k \, \omega} \,\Bigg\{ 1 - \frac{\alpha_s}{4\pi} \Big[C_F-C_A + \mathcal{O}(\epsilon^0) \Big] \Bigg\} \, 2\pi \spac \delta(\nm q - \nm k) + \mathcal{O}(\alpha_s^2) \,,  
    \nonumber\\
    D_2^{(B2)} &= \frac{1}{\nm k \, \omega}\, \Bigg\{ 1 - \frac{\alpha_s}{4\pi} \bigg[ C_F \bigg(\frac{4}{\epsilon} + 5 - 4 \ln \frac{\nm k \, \omega}{-\mu^2} \bigg) - 5 \spac C_A + \mathcal{O}(\epsilon^0) \bigg] \Bigg\} \, 2\pi \spac \delta(\nm q - \nm k)
    \nonumber\\*
    &\quad - \frac{2}{\omega} \, 2\pi \spac \delta'(\nm q - \nm k) + \mathcal{O}(\alpha_s^2) \,.  
\end{align}

\begin{figure}[t]
    \centering
    \includegraphics[scale=0.50]{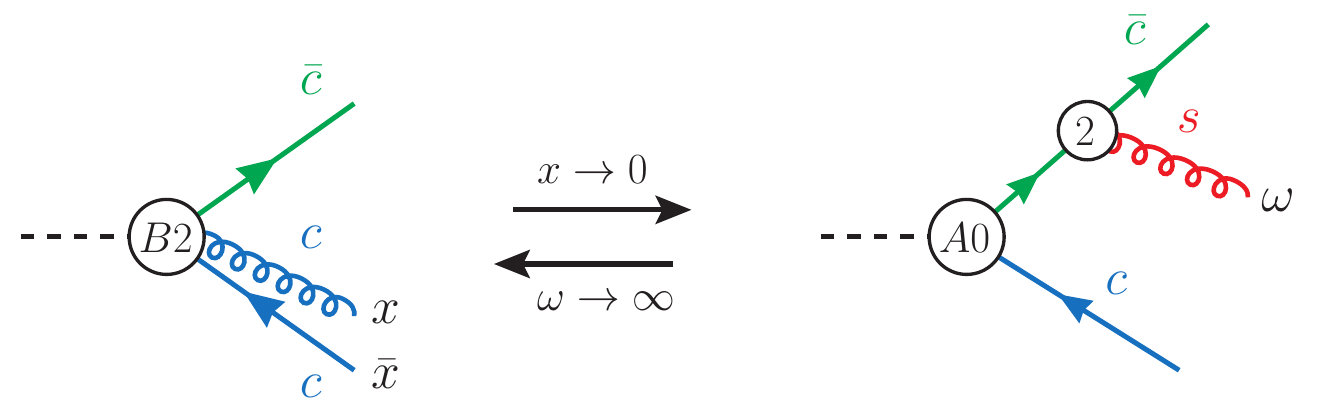}
    \vspace*{0.2cm}
    \caption{Diagrammatic representation of the cancellation of endpoint divergences in~eq.~\eqref{eq:NLP-factorization}. The circled ``2" denotes the $\mathcal{L}_\xi^{(2)}$ insertion. The colours are the same as in fig.~\ref{fig:A-diagrams} and the the soft gluon is drawn in red.}
    \label{fig:CB2-to-DB2}
\end{figure}

In many cases, the convolution in the first line of~eq.~\eqref{eq:NLP-factorization} diverges.
When translated to momentum space, this divergence arises from the limit $x\to0$.
However, at the same time the momentum-space version of the convolution in the second line diverges for $\omega\to\infty$. Physically, this occurs when the emitted collinear gluon in the $(B2)$-type operator becomes soft while the momentum of the soft gluon in the second line becomes collinear, see fig.~\ref{fig:CB2-to-DB2}. In this limit the amplitude in the left figure factorizes over the on-shell singularity of the intermediate anti-collinear outgoing quark propagator as shown on the right, separating the diagram into the leading-power two-jet operator and the soft-collinear splitting  amplitude.
We therefore expect for the $x\to 0$ limit of the eight momentum-space matching coefficients in the first line  of~eq.~\eqref{eq:NLP-factorization} ``refactorization relations'' of the form 
\begin{align} 
    \mathcal{C}_i^{(B2)} \sim \sum_{k} a_{ik} \,C^{(A0)} \otimes D_{k}^{(B2)}  
\end{align}
in terms of suitable linear combinations $a_{ik}$ of the two $D_k^{(B2)}$ functions.
These relations turn the first line of~eq.~\eqref{eq:NLP-factorization} into the form of the second and allow for the cancellation of endpoint divergences.
In the singular limit the on-shell anti-collinear quark propagator propagates only the two physical helicity states of the massless quark.\footnote{See ref.~\cite{Beneke:2022obx} for a related discussion in the context of refactorization of another $C^{(B1)}$ coefficient in the soft-quark limit. } 
Thus only two out of the eight independent operators $\mathcal{J}_i^{(B2)}$ are expected to contribute to endpoint singularities, and therefore all but two short-distance coefficients are finite in the limit $x\to0$ when choosing a convenient basis for the $(B2)$-type operators.

In the basis adopted in this section, the refactorization relations simplify to 
\begin{align} \label{eq:refact-cond}
    \Braces{\mathcal{C}_i^{(B2)}}(\nm k,\np p,x) &= Q^2 \!\int\!\frac{d\nm q}{2\pi} \, C^{(A0)}(\nm q,\np p) \, D_i^{(B2)}(\nm q,\nm k,x \spac \np p) \,, \quad i=1,2 \,,  
    \nonumber\\*
    \Braces{\mathcal{C}_i^{(B2)}}(\nm k,\np p,x) &= 0 \,, \qquad i>2 \,,  
\end{align}
where $\omega \equiv x\spac\np p$ in the argument of the $D^{(B2)}$ functions and 
the notation $\braces{\ldots}$ was introduced in~ref.~\cite{Liu:2019oav} to denote the singular part as $x\to0$.
The factor $Q^2=\nm k \, \np p$ on the right-hand side accounts for the different mass dimension of the operators in the first and second line of~eq.~\eqref{eq:NLP-factorization}.
The right-hand side of~eq.~\eqref{eq:refact-cond} can be computed from~eq.~\eqref{eq:DB2} and the $C^{(A0)}$ coefficient given in~eq.~\eqref{eq:CA0}. 
When performing the $\nm q$ integral in~eq.~\eqref{eq:refact-cond}, the derivative of the delta function in~eq.~\eqref{eq:DB2} translates into a derivative acting on $C^{(A0)}$, and  one finds 
\begin{align}
    \Braces{\mathcal{C}_1^{(B2)}} &= -\frac{1}{2x}\, \bigg\{ 1 + \frac{\alpha_s}{4\pi}  \bigg[C_F \bigg(-\frac{2}{\epsilon^2} + \frac{2 \lnQ - 3}{\epsilon} - \lnsqQ + 3 \lnQ - 9 + \frac{\pi^2}{6} \bigg) + C_A \bigg] + \mathcal{O}(\alpha_s^2) \bigg\} \,,  
    \nonumber\\
    \Braces{\mathcal{C}_2^{(B2)}} &= \frac{1}{x} \,\bigg\{1 + \frac{\alpha_s}{4\pi}  \bigg[C_F \bigg(-\frac{2}{\epsilon^2} + \frac{2 \lnQ - 3}{\epsilon} - \lnsqQ + 3 \spac \lnQ + 4 \Lx - 7 + \frac{\pi^2}{6} \bigg) 
    \nonumber\\*
    &\hspace{25mm} + 5 \spac C_A \bigg] + \mathcal{O}(\alpha_s^2) \bigg\} \,.  
\end{align}
These expressions coincide with the $x \to 0$ expansion of the explicitly computed matching coefficients after converting to $\mathcal{C}_1^{(B2)}$ and $\mathcal{C}_2^{(B2)}$ as detailed in appendix~\ref{app:bases-conv}, thus providing another strong consistency check of our results as well as for the validity of NLP factorization.
We checked that~eq.~\eqref{eq:refact-cond} holds in $d$ dimensions as well.

We note that there are also singularities as $x\to 1$, which correspond to similarly defined splitting functions $D_k^{(B2)}$ of a collinear quark into a collinear gluon and a soft \emph{quark}.
Since the soft quark must interfer with a soft quark from the complex conjugate amplitude, these splitting functions appear only at higher order than $\lambda^2$ in cross sections.
The splitting amplitude into a soft quark at $\mathcal{O}(\lambda)$, $D^{(B1)}$, and its relation to the factorization of the $C^{(B1)}$ coefficient in the soft-quark limit has been discussed in~ref.~\cite{Beneke:2022obx}.


\section{Conclusion}
\label{sec:conclusion}

As the precision of collider measurements continues to improve, the inclusion of next-to-leading-power (NLP) corrections has become increasingly important for extending the predictive power of factorization theorems beyond the leading-power approximation.
These factorization theorems rely on hard matching coefficients of subleading-power operators as essential ingredients. Their determination at the one-loop order is therefore a necessary step toward incorporating NLP hard effects into high-precision resummed and fixed-order predictions.

In this work, we determine the short-distance coefficients of two-jet operators  with up to three fields arising from the matching of (axial-)vector and (pseudo)scalar QCD quark currents on soft-collinear effective theory.
The matching is performed onto subleading-power quark–antiquark operators, both with and without an additional collinear gluon field.
The results are obtained up to $\mathcal{O}(\lambda^2)$ in the power expansion and to next-to-leading order in the strong coupling.
At $\mathcal{O}(\lambda)$, our results are consistent with existing results in refs.~\cite{Strohm:2020,Vladimirov:2021hdn}.
As a first non-trivial check, we confirm that all poles in dimensional regularization are cancelled by the known renormalization factors of the subleading-power operators \cite{Beneke:2017ztn,Beneke:2018rbh}, as required by renormalization-group invariance.
In addition, we verify that the obtained short-distance coefficients satisfy refactorization conditions, ensuring the cancellation of endpoint divergences in subleading-power factorization theorems.

The results obtained in this work provide key ingredients for NLP factorization theorems in processes such as large-$x$ deep-inelastic scattering and $e^+e^- \to 2$ jet production.
A complete phenomenological implementation will also require consistent treatments of subleading-power jet and soft functions.
Together with the present results, this would enable systematic predictions at NLP accuracy for a wide range of collider observables.

\subsubsection*{Acknowledgments}

M.B.~thanks Julian Strohm for collaboration at an earlier stage of the project.
A.R.~thanks Vladyslav Shtabovenko for helpful discussions on the PV reduction of Feynman integrals with vanishing Gram determinants within \textsc{FeynCalc}.
This research was supported by the Excellence Cluster ORIGINS, which is funded by the Deutsche Forschungsgemeinschaft (DFG, German Research Foundation) under Germany’s Excellence Strategy -- EXC 2094 -- 390783311.
AR acknowledges the partial support by the Deutsche Forschungsgemeinschaft (DFG, German Research
Foundation) -- project number 541305755.


\begin{appendix}

\section{Results for the (pseudo)scalar current}
\label{app:scalar-current}

In this appendix, we extend the results from the main text to the cases of a scalar $\bar\psi \psi$ and pseudoscalar $\bar\psi\gamma_5\spac \psi$ QCD current.

\subsubsection*{Operator basis and matching coefficients}

In the scalar case, we choose for the $A$- and $B$-type SCET operators up to $\mathcal{O}(\lambda^2)$
\begin{align}
    J_S^{(A0,A0)}(t,s) &= \bar\Ch_\Cb (t\nm) \, \Ch_\C (s\np) \,,  
    \nonumber\\
    J_{S1}^{(A1,A1)}(t,s) &= \bar\Ch_\Cb(t\nm) \, \frac{i \cev{\slashed{\partial}}_\perp \spac i \slashed{\partial}_\perp}{i \nm \cev{\partial} \, i \np \partial} \, \Ch_\C (s\np) \,,  
    \nonumber\\*
    J_{S2}^{(A1,A1)}(t,s) &= \bar\Ch_\Cb (t\nm) \, \frac{i \cev{\partial}_\perp^\nu \spac i\partial_{\perp\nu}}{i \nm \cev{\partial} \, i \np \partial} \, \Ch_\C (s\np) \,,  
    \nonumber\\
    J_S^{(A0,A2)}(t,s) &= \bar\Ch_\Cb (t\nm) \, \frac{(i\partial_\perp)^2}{i \nm \cev{\partial} \, i \np \partial} \, \Ch_\C (s\np) \,,  
\end{align}
and 
\begin{align} 
    J_{S1}^{(A1,B1)}(t,s_1,s_2) &= \bar\Ch_\Cb (t\nm) \, \frac{i \cev{\slashed{\partial}}_\perp}{i \nm\cev{\partial} \, i\np\partial} \, \As_{\C\perp}(s_1\np) \, \Ch_\C (s_2\np) \,,  
    \nonumber\\
    J_{S2}^{(A1,B1)}(t,s_1,s_2) &= \bar\Ch_\Cb (t\nm) \, \frac{i\cev{\partial}_{\perp\rho}}{i \nm\cev{\partial} \, i \np\partial} \, \A_{\C\perp}^\rho (s_1 n_+) \, \Ch_\C (s_2\np) \,,  
    \nonumber\\
    J_{S1}^{(A0,B2)}(t,s_1,s_2) &= \bar\Ch_\Cb(t\nm) \, \frac{1}{i \nm\cev{\partial} \, i \np\partial} \, i\slashed{\partial}_\perp \big[ \As_{\C\perp}(s_1\np) \, \Ch_\C (s_2\np) \big] \,,  
    \nonumber\\
    J_{S2}^{(A0,B2)}(t,s_1,s_2) &= 
    \bar\Ch_\Cb(t\nm) \, \frac{1}{i \nm\cev{\partial} \, i \np\partial} \, i\partial_{\perp\rho} \big[ \A_{\C\perp}^\rho (s_1\np) \, \Ch_\C (s_2\np) \big] \,,  
    \nonumber\\
    J_{S3}^{(A0,B2)}(t,s_1,s_2) &= \bar\Ch_\Cb(t\nm) \, \frac{1}{i \nm\cev{\partial} \, i \np\partial} \, \As_{\C\perp} (s_1\np) \, i\slashed{\partial}_\perp \Ch_\C (s_2\np) \,,  
    \nonumber\\
    J_{S4}^{(A0,B2)}(t,s_1,s_2) &= \bar\Ch_\Cb(t\nm) \, \frac{1}{i \nm\cev{\partial} \, i \np\partial} \, \A_{\C\perp}^\rho (s_1\np) \, i\partial_{\perp\rho} \spac \Ch_\C (s_2\np) \,,  
\end{align}
where the subscript ``$S$" indicates that these operators are relevant for matching the scalar QCD current.
Note the absence of $\mathcal{O}(\lambda)$ operators in the case of a scalar current.
For the pseudoscalar current $\bar\psi\gamma_5\spac \psi$, one finds $C_i^\text{pseudoscalar} = C_i^\text{scalar}$, when $\gamma_5$ is always placed next to the collinear quark field and when using the naive anti-commuting $\gamma_5$ scheme in dimensional regularization.

The non-vanishing bare short-distance coefficients of the $A$-type operators up to $\mathcal{O}(\lambda^2)$ read 
\begin{align}
    C_S^{(A0)} &= C_{S1}^{(A1,A1)} = 1 + \frac{\alpha_s}{4\pi} \, C_F \bigg(-\frac{2}{\epsilon^2} + \frac{2 \spac \lnQ}{\epsilon} - \lnsqQ - 2  + \frac{\pi^2}{6} \bigg) + \mathcal{O}(\alpha_s^2) \,,  
    \nonumber\\
    C_{S2}^{(A1,A1)} &= 2 \spac \frac{d}{dL} \, C_S^{(A0)} = 
    \frac{\alpha_s}{4\pi} \, 4\spac C_F \bigg(\frac{1}{\epsilon} - \lnQ \bigg) + \mathcal{O}(\alpha_s^2) \,.  
    \nonumber\\[1mm]
    C_S^{(A2)} &= \mathcal{O}(\alpha_s^2) \,.  
\end{align}

There are two independent short-distance coefficients for the $B$-type operators.
We choose
\begin{align}
    C_{S1}^{(B2)} &= -\frac{1}{x} + \frac{\alpha_s}{4\pi} \, C_A \bigg[ \frac{\Lx}{\bar{x}} + \frac{2\spac \Lxb-1}{x} \bigg]
    \nonumber\\*
    &\hspace{12mm} + \frac{\alpha_s}{4\pi} \, C_F \bigg[ \frac{2}{x \spac \epsilon^2} - \frac{2\spac\lnQ}{x \spac \epsilon} - \frac{2\spac\Lx}{\bar{x}} + \frac{\lnsqQ  - 4\spac\Lxb+3}{x} - \frac{\pi^2}{6\spac x} \bigg] + \mathcal{O}(\alpha_s^2) \,,  
    \nonumber\\
    C_{S2}^{(B2)} &= \frac{2}{x} \, \bigg\{ 1 + \frac{\alpha_s}{4\pi} \, C_A \bigg[ 1 - \frac{2\spac\Lxb}{x} \bigg] 
    \nonumber\\*
    &\hspace{14mm} + \frac{\alpha_s}{4\pi} \, C_F \bigg[ -\frac{2}{\epsilon^2} + \frac{2\spac\lnQ}{\epsilon} + \frac{4\Lxb}{x} + \frac{2\spac\Lx}{\bar{x}} - \lnsqQ -1 + \frac{\pi^2}{6} \bigg] \bigg\} + \mathcal{O}(\alpha_s^2) \,,  
\end{align}
where $\lnQ, \Lx$ and $\Lxb$ are defined in~eq.~\eqref{eq:logs}.
The remaining short-distance coefficients are then given by
\begin{align}
    C_{S1}^{(A1,B1)} &= C_S^{(A0)} \,,
    &
    C_{S2}^{(A1,B1)} &= \frac{2}{x} \, C_S^{(A0)} \,,  
    \nonumber\\*
    C_{S3}^{(B2)} &= \frac{1}{\bar{x}} \, C_{S1}^{(B2)} \,,
    &
    C_{S4}^{(B2)} &= - \frac{1}{\bar{x}} \big[ 2 \spac C_{S1}^{(B2)} + C_{S2}^{(B2)} \big] \,.  
\end{align}

\subsubsection*{Renormalization}

The renormalization is similar to the one in the vector case described in sec.~\ref{sec:renormalization}, but as the scalar QCD current is not conserved, the left-hand side of eq.~\eqref{eq:match-rel-mom} contributes a non-trivial $Z$ factor
\begin{align}
    Z_S = 1 + \frac{\alpha_s}{4\pi} \, \frac{3\spac C_F}{\epsilon} + \mathcal{O}(\alpha_s^2) \,.  
\end{align}
The renomalized short-distance coefficients are thus given by
\begin{align} \label{eq:ren-C-scalar}
    C_{Sj}^\mathrm{ren}(\{\underline{x}\}) &= \sum_i \int\!d{\{\underline{y}\}} \,  C_{Si}(\{\underline{y}\}) \, Z_{S,ij}^{-1}(\{\underline{y}\},\{\underline{x}\}) \, Z_S^{-1} \,,  
\end{align}
where the $Z_{S,ij}$ renormalize the SCET operators as in the first line of~eq.~\eqref{eq:ren-J-C}.
The leading-power $Z$ factor is the same as for the vector case, and given in~eq.~\eqref{eq:Z-A0}.
The general structure~\eqref{eq:Z-NNLP} at $\mathcal{O}(\lambda^2)$ again requires the inclusion of two time-ordered product operators.
They are defined as in~eq.~\eqref{eq:J-T1T1} with the replacement $\gamma_\perp^\mu\to1$.
The results in eqs.~\eqref{eq:Z-A2}, \eqref{eq:Z-B2} and~\eqref{eq:Z-A1B1} remain valid.
The entries of the anomalous dimension matrices $\gamma_S^{(B2),(B2)}$ and $\gamma_S^{\A\Ch,\A\Ch}$ are given in appendix~\ref{app:anomalous-dim}.
The mixing of $(A1,A1)$- and $(A1,B1)$-type operators with time-ordered product ones is described by~\cite{Beneke:2018rbh}
\begin{align}
    Z_{S,kj}^{(T1,T1),(A1,A1)} &= \frac{\alpha_s}{\pi} \, \frac{C_F}{\epsilon} \, \delta_{k1} \spac \delta_{j2} + \mathcal{O}(\alpha_s^2) \,,  
    \nonumber\\*[2mm]
    Z_{S,1j}^{(T1,T1),(A1,B1)}(y) &= + \frac{\alpha_s}{\pi} \, \frac{1}{2\epsilon} \bigg(C_F-\frac{C_A}{2}\bigg) \bigg[ -\frac{y}{\bar{y}} \, \delta_{j1} + \frac{2}{\bar{y}} \, \delta_{j2} \bigg] + \mathcal{O}(\alpha_s^2) \,,  
    \nonumber\\*[2mm]
    Z_{S,2j}^{(T1,T1),(A1,B1)}(y) &= - \frac{\alpha_s}{\pi} \, \frac{1}{2\epsilon} \, \frac{C_A}{2} \bigg[ -\delta_{j1} + \frac{2}{y} \, \delta_{j2} \bigg] + \mathcal{O}(\alpha_s^2) \,,  
\end{align}
with $j=1,2$ and $k=1,2$.
We find that eq.~\eqref{eq:ren-C-scalar} removes all poles from the bare short-distance coefficients given above as it should be.

\subsubsection*{Refactorization}

To exhibit the refactorization relations, it is again convenient to change the basis of $(B2)$-type operators to the one given in appendix~\ref{app:bases-conv}.
We then find that~eq.~\eqref{eq:refact-cond} holds for $\mathcal{C}^{(B2)}_{Si}$ in $d$ dimensions with the same $D_k^{(B2)}$ coefficients as given in eq.~\eqref{eq:DB2}.

\section{Hard region of loop integrals}
\label{app:MoR-trick}

When performing matching calculations, it is crucial to only take the hard region of loop integrals into account.
The other regions are reproduced in the EFT by time-ordered products with SCET Lagrangians.

As an example, we consider diagram (b) in fig.~\ref{fig:A-diagrams} which contributes to the matching of the $A$-type operators.
After performing a standard Pasarino-Veltmann (PV) reduction, the two scalar integrals \begin{align} \label{eq:PV-integrals}
    B_0(q^2) &\equiv \tilde{\mu}^{2\epsilon} \! \int\!\frac{d^dl}{(2\pi)^d} \, \frac{1}{l^2 (l - q)^2} \,,
    &
    C_0(2\spac k\!\cdot\!p,p^2) &\equiv \tilde{\mu}^{2\epsilon} \! \int\!\frac{d^dl}{(2\pi)^d} \, \frac{1}{l^2 (l - p)^2 (l + k)^2}   
\end{align}
appear, where $\tilde{\mu}^2\equiv \mu^2e^{\gamma_\mathrm{E}}/4\pi$ and $q\in\{p,p-k\}$.
We assume on shell $k^2=0$ but keep $p^2\neq0$.

We focus on the integral $C_0$, however, the same method applies to $B_0$.
The standard approach is to expand the integrand in the hard loop momentum $l^\mu \sim (1, 1, 1) Q$ prior to integration
\begin{align}
    C_0 &= \tilde{\mu}^{2\epsilon} \int\!\frac{d^dl}{(2\pi)^d} \, \frac{1}{l^2 (l - p_-)^2 (l + k_+)^2} \bigg[ 1 + \underbrace{\frac{2\spac l\!\cdot\!p_\perp}{(l-p_-)^2}}_{=0} - \underbrace{\frac{2\spac l\!\cdot\!k_\perp}{(l+k_+)^2}}_{=0} + \frac{2\spac l\!\cdot\!p_+}{(l-p_-)^2} - \frac{2\spac l\!\cdot\!k_-}{(l+k_+)^2}
    \nonumber\\*
    &\quad + \frac{(2\spac l\!\cdot\!p_\perp)^2}{(l-p_-)^4} - \frac{(2\spac l\!\cdot\!k_\perp)^2}{(l+k_+)^4} - \frac{(2\spac l\!\cdot\!p_\perp) (2\spac l\!\cdot\!k_\perp)}{(l-p_-)^2(l+k_+)^2} - \frac{p^2}{(l-p_-)^2} - \frac{k^2}{(l+k_+)^2} + \mathcal{O}(\lambda^3) \bigg] \,,  
\end{align}
where $q_\pm^\mu=n_\mp  q\, \frac{n_\pm^\mu}{2}$ and one should think of $\np k$ inside $k_-$ as being eliminated using $k^2=0$.
The two $\mathcal{O}(\lambda)$ terms vanish because the denominators do not contain transverse momentum components.
The remaining vector and tensor integrals can again be reduced to scalar integrals applying PV reduction.
Using this method, one has to calculate several integrals with different propagator powers.
This is inconvenient for more complicated diagrams, such as those shown in fig.~\ref{fig:B-diagrams}.

Instead of expanding the Feynman integrals in the representation~\eqref{eq:PV-integrals}, one can first introduce Feynman parameters to write
\begin{align}
    C_0 = \tilde{\mu}^{2\epsilon} \int_0^1\!d\beta_1 \! \int_0^{1-\beta_1} \! d\beta_2 \! \int \! \frac{d^dl}{(2\pi)^d} \, \frac{2}{\left[(l-P)^2 + \Delta\right]^3} \,,  
\end{align}
where for on shell $k^2 = 0$
\begin{align}
    P^\mu &= \beta_1 \spac p^\mu - \beta_2 \spac k^\mu \,,
    &
    \Delta &= \beta_1\beta_2 \, 2 \spac p\!\cdot\! k + \beta_1(1-\beta_1) \, p^2 \,.   
\end{align}
In general, in a method-of-region calculation, shifting the loop momentum by a hard momentum $P^\mu\sim(1,1,\lambda)Q$ before expansion changes its scaling and is only allowed if one intends to expand in the hard region.
Hence the shift $l\to l-P$ is perfectly fine for a matching calculation.
We find
\begin{align}
    C_0 = \tilde{\mu}^{2\epsilon} \int_0^1\!d\beta_1 \! \int_0^{1-\beta_1} \! d\beta_2 \! \int \! \frac{d^dl}{(2\pi)^d} \, \frac{2}{\left[l^2 + \Delta\right]^3} \,.  
\end{align}
The expansion in $\lambda$ arises from
\begin{align}
    \Delta &= \beta_1\beta_2 \, Q^2 + \beta_1\beta_2 \, 2 \spac p_\perp\!\cdot\!k_\perp + \beta_1(1-\beta_1) \, p^2 + \mathcal{O}(\lambda^4) \,,  
\end{align}
where the second and third term are of $\mathcal{O}(\lambda^2)$.
The hard-region expansion of the integral then reads
\begin{align}
    C_0 &= \tilde{\mu}^{2\epsilon} \int_0^1\!d\beta_1 \! \int_0^{1-\beta_1} \! d\beta_2 \! \int \! \frac{d^dl}{(2\pi)^d} \, \frac{2}{\left[l^2 + \beta_1\beta_2 \, Q^2\right]^3}
    \nonumber\\*
    &\hspace{20mm} \times \bigg[ 1 - 3\,\frac{\beta_1\beta_2 \, 2 \spac p_\perp\!\cdot\!k_\perp + \beta_1(1-\beta_1) \, p^2}{l^2 + \beta_1\beta_2 \, Q^2} + \mathcal{O}(\lambda^4) \bigg]  
    \\[2mm]\nonumber
    &= \frac{i}{16\pi^2} \, \frac{1}{Q^2} \bigg(\frac{-Q^2}{\mu^2}\bigg)^{\!-\epsilon} \, \frac{e^{\epsilon\gamma_\mathrm{E}} \, \Gamma^2(-\epsilon) \, \Gamma(1+\epsilon)}{\Gamma(1-2\epsilon)} \bigg[ 1 - (1+\epsilon) \, \frac{2 \spac p_\perp\!\cdot\!k_\perp}{Q^2} - \epsilon \, \frac{p^2}{Q^2} + \mathcal{O}(\lambda^4) \bigg] \,.  
\end{align}
Performing a similar calculation for the integral $B_0$, we find that $\Lambda^\mu$ defined in~eq.~\eqref{eq:B-onshell-diag} has the general form
\begin{align} \label{eq:Lam-FFs}
    \Lambda^\mu (p^2, p\!\cdot\!k) &= \bigg[f_0^{(0)}(Q^2) + f_0^{(1)}(Q^2) \, \frac{2 \spac p_\perp\!\cdot\!k_\perp}{Q^2} + f_0^{(2)}(Q^2) \, \frac{p^2}{Q^2} \bigg] \,\gamma^\mu 
    \nonumber\\*
    &+ \bigg[f_1^{(0)}(Q^2) + f_1^{(1)}(Q^2) \, \frac{2 \spac p_\perp\!\cdot\!k_\perp}{Q^2} + f_1^{(2)}(Q^2) \, \frac{p^2}{Q^2} \bigg]\, \slashed{p} \, p^\mu
    \nonumber\\*
    &+ \bigg[f_2^{(0)}(Q^2) + f_2^{(1)}(Q^2) \, \frac{2 \spac p_\perp\!\cdot\!k_\perp}{Q^2} + f_2^{(2)}(Q^2) \, \frac{p^2}{Q^2} \bigg] \,\slashed{p} \, k^\mu + \mathcal{O}(\lambda^4) \,,   
\end{align}
where we used $\bar{u}(k) \spac \slashed{k} = 0$, since the outgoing quark  is on-shell.
The functions $f_i^{(j)}$ can be identified with form factors, which are already expanded in $\lambda$.
We find
\begin{align}
    f_0^{(0)} &= C^{(A0)} \,, \qquad \qquad
    f_0^{(1)} = \frac{1}{2} \, C_2^{(A1,A1)} = \frac{d}{dL} \spac C^{(A0)} \,,  
    \nonumber\\
    f_0^{(2)} &= - \frac{\alpha_s}{4\pi} \, C_F \bigg[\frac{2}{\epsilon^2} + \frac{1}{\epsilon} \, (1 - 2 \lnQ ) + \lnsqQ - \lnQ + 5 - \frac{\pi^2}{6} \bigg] + \mathcal{O}(\alpha_s^2) \,,  
    \nonumber\\
    f_1^{(0)} &= - \frac{\alpha_s}{4\pi} \, \frac{2\spac C_F}{Q^2} \bigg[ \frac{1}{\epsilon} - \lnQ + 1 \bigg] + \mathcal{O}(\alpha_s^2) \,,  
    \nonumber\\
    f_2^{(0)} &= \frac{\alpha_s}{4\pi} \, \frac{2\spac C_F}{Q^2} \bigg[\frac{2}{\epsilon^2} - \frac{2}{\epsilon} \, (\lnQ - 2) + \lnsqQ - 4 \lnQ + 9 - \frac{\pi^2}{6} \bigg] + \mathcal{O}(\alpha_s^2) \,,  
\end{align}
with $C^{(A0)}$ given in eq.~\eqref{eq:CA0}.

We can use these expressions to extract the ``local'' part of the one-particle-reducible diagrams of fig.~\ref{fig:B-diagrams-onshell}, which 
belongs to the $B$-type operator matching coefficients.
It is evident that the factor $p^2$ multiplying $f_0^{(2)}$ in~eq.~\eqref{eq:Lam-FFs} can cancel the $1/p^2$ in~eq.~\eqref{eq:propagator_decomp} yielding a local term contributing to the matching coefficient.
Note, however, that the terms in~eq.~\eqref{eq:Lam-FFs} proportional to $\slashed p$ combine with the quark-propagator in~eq.~\eqref{eq:B-onshell-diag}, ``hitting" the $1/p^2$ pole and thus yielding additional local contributions given in terms of the form-factors $f_1^{(0)}$ and $f_2^{(0)}$.
Consequently, the functions $f_{1,2}^{(1)}$ and $f_{1,2}^{(2)}$ yield corrections of order $\mathcal{O}(\lambda^3)$, i.e.~beyond the accuracy of the current analysis.
For this reason we do not display their expressions here.
Afterwards, to obtain the contribution to the different $B$-type matching coefficients, one also needs to expand the Dirac structures together with the remaining terms on the right-hand side of~eq.~\eqref{eq:B-onshell-diag}, in the same way as in the computation of the diagrams in fig.~\ref{fig:B-diagrams}.
We checked that the remaining, non-local terms proportional to $1/p^2$ are reproduced -- order-by-order in the $\lambda$ expansion -- in SCET by time-ordered products.


\section{Off-shell matching}
\label{app:off-shell-match}

In the main text, we described how to obtain the short-distance coefficients from on-shell matching.
In this case, one has to include one-particle-reducible diagrams such as the ones shown in fig.~\ref{fig:B-diagrams-onshell}.
If one instead matches \emph{off-shell}, one only needs to consider 1PI diagrams and the missing terms are obtained by applying the equations of motion at operator level. Off-shell, one \emph{cannot} use $p^2=k^2=0$ and the relations~\eqref{eq:QCD-SCET-spinor} for the spinors. 
Instead, the small components of the (anti-)collinear quark fields $\eta_{\C},(\eta_{\Cb})$, which are removed in the SCET construction using the equations of motion, must be kept as building blocks for operators.\footnote{This is analogous to the gluon field component $\nm\A_\C$. 
However, in this case one usually keeps $\nm\A_\C$ in the SCET construction, even though one could remove it by means of~eq.~\eqref{eq:gluon-eom} below analogously to removing $\eta_\C$.} 
They fulfil $\nps\eta_\C=\nms\eta_\Cb=0$ and are of $\mathcal{O}(\lambda^2)$ in the power expansion.

The $\eta_\C$ field components counts like a transverse derivative and provides a factor of $\lambda$ suppression relative to the large spinor component $\Ch_\C = W_\C^\dagger\xi_\C$. One thus has to include two additional $(A1)$-type operators at $\mathcal{O}(\lambda)$,
\begin{align} \label{eq:EOM-op-1}
    \tilde{J}_1^{(A0,A1)}(t,s) &= \bar\Ch_\Cb(t\nm) \, \np^\mu \, \frac{\nms}{2} \, (W_\C^\dagger \eta_\C)(s\np) \,,  
    \nonumber\\*
    \tilde{J}_2^{(A0,A1)}(t,s) &= \bar\Ch_\Cb(t\nm) \, \nm^\mu \, \frac{\nms}{2} \, \frac{i\np\partial}{i\nm\cev{\partial}} \, (W_\C^\dagger \eta_\C)(s\np) \,.  
\end{align}
At $\mathcal{O}(\lambda^2)$, one needs five additional operators of $(A1,A1)$ type,
\begin{align}
    \tilde{J}_1^{(A1,A1)}(t,s) &= (\bar\eta_\Cb \, W_\Cb)(t\nm) \, \gamma_\perp^\mu \, (W_\C^\dagger \eta_\C)(s\np) \,,  
    \nonumber\\
    \tilde{J}_2^{(A1,A1)}(t,s) &= \bar\Ch_\Cb(t\nm) \, \frac{i\cev{\partial}_\perp^\mu}{i\nm\cev{\partial}} \, \frac{\nms}{2} \, (W_\C^\dagger \eta_\C)(s\np) \,,  
    \nonumber\\[-1mm]
    \tilde{J}_3^{(A1,A1)}(t,s) &= (\bar\eta_\Cb \, W_\Cb)(t\nm) \, \frac{\nps}{2} \, \frac{i\partial_\perp^\mu}{i\np\partial} \, \Ch_\C(s\np) \,,  
    \nonumber\\[-1mm]
    \tilde{J}_4^{(A1,A1)}(t,s) &= \bar\Ch_\Cb(t\nm) \, \frac{i\cev{\slashed{\partial}}_\perp}{i\nm\cev{\partial}} \, \gamma_\perp^\mu \, \frac{\nms}{2} \, (W_\C^\dagger \eta_\C)(s\np) \,,  
    \nonumber\\[-1mm]
    \tilde{J}_5^{(A1,A1)}(t,s) &= (\bar\eta_\Cb \, W_\Cb)(t\nm) \, \frac{\nps}{2} \, \gamma_\perp^\mu \, \frac{i\slashed{\partial}_\perp}{i\np\partial} \, \Ch_\C(s\np) \,,  
\end{align}
and two of the $(A2)$ type,
\begin{align} \label{eq:EOM-op-3}
    \tilde{J}_1^{(A0,A2)}(t,s) &= \bar\Ch_\Cb(t\nm) \, \gamma_\perp^\mu \, \frac{i\nm D_s}{i\nm\cev{\partial}} \, \Ch_\C(s\np) \,,  
    \nonumber\\*
    \tilde{J}_2^{(A0,A2)}(t,s) &= \bar\Ch_\Cb(t\nm) \, \frac{i\partial_\perp^\mu}{i\nm\cev{\partial}} \, \frac{\nms}{2} \, (W_\C^\dagger \eta_\C)(s\np) \,.  
\end{align}
As always, there are similar operators with $\C\leftrightarrow\Cb$ and $\np\leftrightarrow\nm$.

The operators in~eqs.~\eqref{eq:EOM-op-1} to~\eqref{eq:EOM-op-3} are related to the basis defined in sec.~\ref{sec:basis} by the operator equations of motion for the collinear quark field~\cite{Beneke:2002ph}
\begin{align}
    W_\C^\dagger \eta_\C &= - \frac{1}{i\np\partial} \, \frac{\nps}{2} \, \big(i\slashed{\partial}_\perp + \As_{\C\perp} \big) \Ch_\C \,,   
    \nonumber\\*
    i\nm D_s \, \Ch_\C &= - \Big[\nm\A_\C + \big(i\slashed{\partial}_\perp + \As_{\C\perp}\big) \frac{1}{i\np\partial} \big(i\slashed{\partial}_\perp + \As_{\C\perp}\big) \Big] \Ch_\C \,.  
\end{align}
and the collinear gluon field~\cite{Beneke:2017ztn}
\begin{align} \label{eq:gluon-eom}
    \nm\A_\C = - \frac{2}{i\np\partial} \, i\partial_\perp^\nu \, \A_{\C\perp,\nu} - \frac{2}{(i\np\partial)^2} \, \big[\A_{\C\perp}^\nu , (i\np\partial \, \A_{\C\perp,\nu}) \big] - \frac{4g_s^2}{(i\np\partial)^2} \, t^a \spac \big[\Ch_\C \, t^a \, \frac{\nps}{2} \Ch_\C\big] \,.  
\end{align}
For the present analysis, one can ignore all soft fields except the one in $\nm D_s$ as they start contributing at $\mathcal{O}(\lambda^3)$ only.
One finds for example
\begin{align}
    \tilde{J}_1^{(A1)}(t,s) &= - J_i^{(A1)}(t,s) - J_i^{(B1)}(t,s,s) \,, \qquad i=1,2 \,,  
\end{align}
which means that the short-distance coefficients of $J_i^{(A1)}$ and $J_i^{(B1)}$ are modified as $C_i^{(A1)}\to C_i^{(A1)}-\tilde{C}_i^{(A1)}$ and $C_i^{(B1)}\to C_i^{(B1)}-\tilde{C}_i^{(A1)}$ for $i=1,2$ when applying the  equations of motion to eliminate the off-shell operators.
We performed both on-shell and off-shell matching and checked that the results agree.

\section{Anomalous dimensions}
\label{app:anomalous-dim}

In this appendix, we present the anomalous dimensions matrices of the $(A1,B1)$ and $(A0,B2)$-type operators.
They can also be found in an ancillary \texttt{Mathematica} notebook.

\subsubsection*{Vector current}

We find that the $8 \times 8$ anomalous dimension matrix $\gamma^{(B2),(B2)}$ in~eq.~\eqref{eq:gamma_B2_split} has 30 
non-vanishing entries within our basis convention in sec.~\ref{sec:basis}.
However, because of the evanescent operators in eq.~\eqref{eq:evanescent_operators} and RPI relations, the elements of this matrix can be expressed as linear combination of  fewer ``basis" elements.
To achieve a maximal reduction to ``independent" entries, one needs to allow the coefficients to be functions of $x,y$.
The explicit expression are not particularly illuminating, so we do not show them here. All entries can be found in the ancillary \texttt{Mathematica} notebook.
As an example, the entries of the upper left block $\gamma^{\partial[\A\Ch],\partial[\A\Ch]}$ in~eq.~\eqref{eq:gamma_B2_split} fulfil
\begin{align}
    \gamma^{\partial[\A\Ch],\partial[\A\Ch]}_{11} &= \gamma^{\partial[\A\Ch],\partial[\A\Ch]}_{33} = \gamma^{\A\Ch,\A\Ch}\,,  
    \nonumber\\
    \gamma^{\partial[\A\Ch],\partial[\A\Ch]}_{41} &= \frac{1}{2}\spac \gamma^{\partial[\A\Ch],\partial[\A\Ch]}_{23} = - \gamma^{\partial[\A\Ch],\partial[\A\Ch]}_{21}\,,   
    \nonumber\\
    \gamma^{\partial[\A\Ch],\partial[\A\Ch]}_{22} &= \gamma^{\partial[\A\Ch],\partial[\A\Ch]}_{44} 
    = \gamma^{\A\Ch,\A\Ch} + 2 \spac \gamma^{\partial[\A\Ch],\partial[\A\Ch]}_{21}\,,  
\end{align}
where $\gamma^{\A\Ch,\A\Ch}$ is given in~eq.~\eqref{eq:gamma-ACh-ACh} and
\begin{align}
    \gamma^{\partial[\A\Ch],\partial[\A\Ch]}_{21} &= \frac{\alpha_s}{4\pi} \,\frac{C_A}{2} \bigg\{ \frac{\theta(x-y) \spac \bar{x}}{x \bar{y}} + \frac{\theta(y-x)}{y} \bigg\} \, (x+y) - \frac{\alpha_s}{4\pi} \, C_F \, \bar{x}
    \nonumber\\*
    &\phantom{=} - \frac{\alpha_s}{4\pi} \bigg( C_F  - \frac{C_A}{2} \bigg)
    \bigg\{ \frac{\theta(x - \bar{y}) \spac \bar{x}}{x y} \, (x + y) + \frac{\theta(\bar{y} - x)}{\bar{y}} \, (x - \bar{y}) \bigg\} + \mathcal{O}(\alpha_s^2) \,.   
\end{align}

The anomalous dimension matrix $\gamma_{ij}^{\A\Ch,\A\Ch}$ in~eq.~\eqref{eq:Z-A1B1} of the $(A1,B1)$-type operators has only seven non-zero entries.
The only (new) independent entry is
\begin{align}
    \gamma_{22}^{\A\Ch,\A\Ch} = \gamma_{44}^{\A\Ch,\A\Ch} ={} & - \frac{\alpha_s}{4\pi} \, C_A \bigg\{ \bigg[\frac{\theta(x-y)}{x-y}\bigg]_+ + \bigg[\frac{\theta(y-x)}{y-x}\bigg]_+ - \frac{\theta(x-y)}{x\bar{y}} - \frac{\theta(y-x)}{y} \bigg\}
    \nonumber\\*
    &-\frac{\alpha_s}{2\pi} \bigg( C_F - \frac{C_A}{2} \bigg) \frac{\theta(x-\bar{y}) \spac \bar{x}}{xy} + \mathcal{O}(\alpha_s^2) \,.  
\end{align}
The remaining entries are given by
\begin{align}
    \gamma_{11}^{\A\Ch,\A\Ch} &= \gamma_{33}^{\A\Ch,\A\Ch} = \gamma^{\A\Ch,\A\Ch} \,,  
    \nonumber\\*
    2 \gamma_{21}^{\A\Ch,\A\Ch} &= - 2 \spac \gamma_{41}^{\A\Ch,\A\Ch} = \spac \gamma_{43}^{\A\Ch,\A\Ch} = \gamma^{\A\Ch,\A\Ch} - \gamma_{22}^{\A\Ch,\A\Ch} \,,  
\end{align}
where $\gamma^{\A\Ch,\A\Ch}$ is given in~eq.~\eqref{eq:gamma-ACh-ACh}.

\subsubsection*{Scalar current}

For the scalar current, each non-trivial block in~eq.~\eqref{eq:gamma_B2_split} has only two independent entries as the ten non-zero entries fulfil
\begin{align}
     \gamma_{S,11}^{\partial[\A\Ch],\partial[\A\Ch]} &= 2\spac \gamma_{S,21}^{\partial[\A\Ch],\partial[\A\Ch]} + \gamma_{S,22}^{\partial[\A\Ch],\partial[\A\Ch]} \,,  
    \nonumber\\*
    \gamma_{S,44}^{\A\partial\Ch,\A\partial\Ch} &= \gamma_{S,33}^{\A\partial\Ch,\A\partial\Ch} - 2\spac \gamma_{S,43}^{\A\partial\Ch,\A\partial\Ch} \,,   
    \nonumber\\*
    \gamma_{S,32}^{\A\partial\Ch,\partial[\A\Ch]} &= - 2 \spac \gamma_{S,31}^{\A\partial\Ch,\partial[\A\Ch]} = 2 \spac \gamma_{S,42}^{\A\partial\Ch,\partial[\A\Ch]} \,,  
\end{align}
where the numerical subscript refers to the entry of the original $\gamma^{(B2),(B2)}$ matrix. 
We find
\begin{align} \label{eq:gamma-ACh-ACh-scalar}
    \gamma_{S,21}^{\partial[\A\Ch],\partial[\A\Ch]} ={}& -\frac{\alpha_s}{4\pi} \, \frac{C_A}{2}\, \bigg\{ \frac{\theta(x-y) \spac \bar{x}}{x\bar{y}} + \frac{\theta(y-x)}{y} \bigg\} \, (x+y) + \frac{\alpha_s}{4\pi} \, C_F \, \bar{x}
    \nonumber\\*
    &+ \frac{\alpha_s}{4\pi} \,\bigg(C_F-\frac{C_A}{2}\bigg) \bigg\{ \frac{\theta(x-\bar{y}) \spac \bar{x}}{xy} \, (x+y) - \frac{\theta(\bar{y}-x)}{\bar{y}} \, (\bar{x}-y) \bigg\} + \mathcal{O}(\alpha_s^2) \,,  
    \nonumber\\
    \gamma_{S,22}^{\partial[\A\Ch],\partial[\A\Ch]} ={}& -\frac{\alpha_s}{4\pi} \, C_A\, \bigg\{ \bigg[\frac{\theta(x-y)}{x-y}\bigg]_+ + \bigg[\frac{\theta(y-x)}{y-x}\bigg]_+ - \frac{\theta(x-y)}{x\bar{y}} - \frac{\theta(y-x)}{y} \bigg\}
    \nonumber\\*
    &- \frac{\alpha_s}{2\pi} \,\bigg(C_F-\frac{C_A}{2}\bigg) \, \frac{\theta(x-\bar{y}) \spac \bar{x}}{xy} + \mathcal{O}(\alpha_s^2) \,,  
\end{align}
and
\begin{align}
    \gamma_{S,33}^{\A\partial\Ch,\A\partial\Ch} ={}& -\frac{\alpha_s}{4\pi} \, C_A \bigg\{ \bigg[\frac{\theta(x-y)}{x-y}\bigg]_+ + \bigg[\frac{\theta(y-x)}{y-x}\bigg]_+ - \frac{\theta(x-y) \spac x}{\bar{y}} + \frac{\theta(y-x)}{y\bar{y}} \, (y-\bar{x}\bar{y}) \bigg\} 
    \nonumber\\*
    &+ \frac{\alpha_s}{2\pi} \,\bigg(C_F-\frac{C_A}{2}\bigg) \bigg\{ \frac{\theta(x-\bar{y})}{y} - \frac{\theta(\bar{y}-x)}{\bar{y}} \bigg\} \, \bar{x} + \frac{\alpha_s}{2\pi} \, C_F \, \frac{\bar{x}}{\bar{y}} + \mathcal{O}(\alpha_s^2) \,,  
    \nonumber\\
    \gamma_{S,43}^{\A\partial\Ch,\A\partial\Ch} ={}& - \frac{\alpha_s}{4\pi} \, \frac{C_A}{2} \bigg\{ \frac{\theta(x-y) \spac \bar{x}}{\bar{y}^2} \, (x+\bar{y}) - \frac{\theta(y-x)}{y\bar{y}} \, \big[x\bar{x} -y(2-x)\big] \bigg\} + \frac{\alpha_s}{4\pi} \, C_F \, \frac{\bar{x}\spac (2-x)}{\bar{y}}
    \nonumber\\*
    &+ \frac{\alpha_s}{4\pi} \,\bigg(C_F-\frac{C_A}{2}\bigg) \bigg\{ \frac{\theta(x-\bar{y}) \spac \bar{x}}{y\bar{y}} \, (x-y) - \frac{\theta(\bar{y}-x)}{\bar{y}^2} \, \big[2\bar{y}-x(x+\bar{y})\big] \bigg\} + \mathcal{O}(\alpha_s^2) \,,   
\end{align}
and
\begin{align}
    \gamma_{S,31}^{\A\partial\Ch,\partial[\A\Ch]} ={}& - \frac{\alpha_s}{4\pi} \, C_A \bigg\{ \frac{\theta(x-y)}{x\bar{y}} \, (1+x\bar{y}) + \theta(y-x) \bigg\} \, \bar{x} + \frac{\alpha_s}{2\pi} \, C_F \, \bar{x}
    \nonumber\\*
    &+ \frac{\alpha_s}{2\pi} \,\bigg(C_F-\frac{C_A}{2}\bigg) \,\bigg\{ \frac{\theta(x-\bar{y})}{xy} \, (1-xy) - \theta(\bar{y}-x) \bigg\} \, \bar{x} + \mathcal{O}(\alpha_s^2) \,,  
    \nonumber\\ 
    \gamma_{S,41}^{\A\partial\Ch,\partial[\A\Ch]} ={}& -\frac{\alpha_s}{4\pi} \, \frac{C_A}{2} \bigg\{ \frac{\theta(x-y) \spac \bar{x}}{x\bar{y}} \, (y+2x\bar{y}) + \theta(y-x) \spac (1+2\bar{x}) \bigg\} + \frac{\alpha_s}{4\pi} \, C_F \, \bar{x} \spac (1+2\bar{x})
    \nonumber\\*
    &+ \frac{\alpha_s}{4\pi} \,\bigg(C_F-\frac{C_A}{2}\bigg) \,\bigg\{ \frac{\theta(x-\bar{y}) \spac \bar{x}}{xy} \, (x\bar{y}+\bar{x}y) - \frac{\theta(\bar{y}-x)}{\bar{y}} \, (2 \bar{x} \bar{y} + \bar{x} - y) \bigg\}
    \nonumber\\*[1mm]
    &+ \mathcal{O}(\alpha_s^2) \,,  
\end{align}
The three non-zero entries of the anomalous dimension matrix $\gamma_{S,ij}^{\A\Ch,\A\Ch}$ of the $(A1,B1)$-type operators fulfil
\begin{align}
    \gamma_{S,21}^{\A\Ch,\A\Ch} &= \gamma_{S,21}^{\partial[\A\Ch],\partial[\A\Ch]} \,,  
    \nonumber\\
    \gamma_{S,22}^{\A\Ch,\A\Ch} &= \gamma_{S,22}^{\partial[\A\Ch],\partial[\A\Ch]} \,,  
    \nonumber\\
    \gamma_{S,11}^{\A\Ch,\A\Ch} &= 2\spac \gamma_{S,21}^{\A\Ch,\A\Ch} + \gamma_{S,22}^{\A\Ch,\A\Ch} \,,  
\end{align}
with the anomalous dimensions given in~eq.~\eqref{eq:gamma-ACh-ACh-scalar}.

\section{Alternative \texorpdfstring{\boldmath $(A0,B2)$}{(A0,B2)} basis}
\label{app:bases-conv}

We introduce an alternative basis of the $(A0,B2)$-type operators, in which the transverse derivatives act on the collinear quark and gluon fields separately.
This choice is convenient for the refactorization discussion in sec.~\ref{sec:refactorization} and appendix~\ref{app:scalar-current}.

\subsubsection*{Vector current}
For the vector current, the operators in the alternative basis are given by
\begin{align}
    \mathcal{J}_1^{(A0,B2)} (t,s_1,s_2) &= \bar\Ch_\Cb(t\nm) \, \frac{1}{i \nm\cev{\partial} \, i \np\partial} \big[i\partial_\perp^\rho \A_{\C\perp}^\sigma (s_1\np) \big] [\gamma_{\perp\rho} \spac , \spac \gamma_{\perp\sigma}] \gamma_\perp^\mu \, \Ch_\C (s_2\np) \,,  
    \nonumber\\
    \mathcal{J}_2^{(A0,B2)} (t,s_1,s_2) &= \bar\Ch_\Cb(t\nm) \, \frac{\gamma_\perp^\mu}{i \nm\cev{\partial} \, i \np\partial} \big[ i\partial_{\perp\rho} \, \A_{\C\perp}^\rho (s_1\np) \big] \Ch_\C (s_2\np) \,,  
    \nonumber\\
    \mathcal{J}_3^{(A0,B2)} (t,s_1,s_2) &= \bar\Ch_\Cb(t\nm) \, \frac{1}{i \nm\cev{\partial} \, i \np\partial} \big[i\partial_\perp^\mu \As_{\C\perp} (s_1\np) \big]  \Ch_\C (s_2\np) \,,  
    \nonumber\\
    \mathcal{J}_4^{(A0,B2)} (t,s_1,s_2) &= \bar\Ch_\Cb(t\nm) \, \frac{1}{i \nm\cev{\partial} \, i \np\partial} \big[i\slashed{\partial}_\perp \A_{\C\perp}^\mu (s_1\np)\big] \Ch_\C (s_2\np) \,,  
    \nonumber\\
    \mathcal{J}_i^{(A0,B2)} (t,s_1,s_2) &= J_i^{(A0,B2)} (t,s_1,s_2) \,, \qquad i=5,6,7,8  \,,  
\end{align}
and the corresponding short-distance coefficients are denoted by $\mathcal{C}_i^{(B2)}$.
They are related to the ones of the operators defined in sec.~\ref{sec:basis} by
\begin{align}
    \mathcal{C}_1^{(B2)} &= \frac{1}{2} \, C_1^{(B2)} \,,
    &
    \mathcal{C}_2^{(B2)} &= C_4^{(B2)} + C_1^{(B2)} \,,  
    \nonumber\\
    \mathcal{C}_3^{(B2)} &= C_3^{(B2)} + 2 \spac C_1^{(B2)} \,,
    &
    \mathcal{C}_4^{(B2)} &= C_2^{(B2)} - 2 \spac C_1^{(B2)} \,,  
    \nonumber\\
    \mathcal{C}_5^{(B2)} &= C_5^{(B2)} - C_1^{(B2)} \,,
    &
    \mathcal{C}_6^{(B2)} &= C_6^{(B2)} + C_2^{(B2)} \,,  
    \nonumber\\
    \mathcal{C}_7^{(B2)} &= C_7^{(B2)} + C_3^{(B2)} \,,
    &
    \mathcal{C}_8^{(B2)} &= C_8^{(B2)} + C_4^{(B2)} + 2 \spac C_1^{(B2)} \,.  
\end{align}

\subsubsection*{Scalar current}
For the scalar current, we use as alternative basis
\begin{align}
    \mathcal{J}_{S1}^{(A0,B2)} (t,s_1,s_2) &= \bar\Ch_\Cb(t\nm) \, \frac{1}{i \nm\cev{\partial} \, i \np\partial} \big[i\partial_\perp^\rho \A_{\C\perp}^\sigma (s_1\np) \big] [\gamma_{\perp\rho} \spac , \spac \gamma_{\perp\sigma}] \spac \Ch_\C (s_2\np) \,,  
    \nonumber\\
    \mathcal{J}_{S2}^{(A0,B2)} (t,s_1,s_2) &= \bar\Ch_\Cb(t\nm) \, \frac{1}{i \nm\cev{\partial} \, i \np\partial} \big[ i\partial_{\perp\rho} \, \A_{\C\perp}^\rho (s_1\np) \big] \spac \Ch_\C (s_2\np) \,,  
    \nonumber\\
    \mathcal{J}_{Si}^{(A0,B2)} (t,s_1,s_2) &= J_{Si}^{(A0,B2)} (t,s_1,s_2) \,, \qquad i=3,4 \,.  
\end{align}
Their short-distance coefficients $\mathcal{C}_{Si}^{(B2)}$ are related to the ones given in appendix~\ref{app:scalar-current} by
\begin{align}
    \mathcal{C}_{S1}^{(B2)} &= \frac{1}{2} \, C_{S1}^{(B2)} \,,
    &
    \mathcal{C}_{S2}^{(B2)} &= C_{S2}^{(B2)} + C_{S1}^{(B2)} \,,   
    \nonumber\\
    \mathcal{C}_{S3}^{(B2)} &= C_{S3}^{(B2)} - C_{S1}^{(B2)} \,,
    &
    \mathcal{C}_{S4}^{(B2)} &=  C_{S4}^{(B2)}  + C_{S2}^{(B2)} + 2 \spac C_{S1}^{(B2)} \,.  
\end{align}

\end{appendix}

\pdfbookmark[1]{References}{Refs}
\bibliography{references.bib}

\end{document}